\documentclass[pra,reprint,showpacs,superscriptaddress,twocolumn]{revtex4-2}

\usepackage[utf8]{inputenc}      
\usepackage[T1]{fontenc}         
\usepackage{times}               

\usepackage{graphicx}            
\usepackage{dcolumn}             
\usepackage{multirow}            
\usepackage{xcolor}              
\usepackage{lipsum}
\usepackage{booktabs}  
\usepackage{amsmath, amssymb, amsthm, mathtools, bm, bbm, dsfont}

\usepackage{empheq, cases}       

\usepackage[colorlinks=true,
            linkcolor=blue!50!black,
            citecolor=red,
            urlcolor=blue,
            pdfauthor={},
            pdftitle={},
            pdfsubject={},
            pdfkeywords={}]{hyperref}

\usepackage{tikz}
\usepackage{pgfplots}
\usepackage{tikz-3dplot}
\pgfplotsset{compat=newest}
\usepgfplotslibrary{colormaps, patchplots}
\usetikzlibrary{arrows.meta, decorations.text, decorations.markings,
                shapes.symbols, shapes.arrows, intersections,
                mindmap, trees, backgrounds, fadings, shadows,
                positioning, calc, patterns, snakes}

\definecolor{c1}{RGB}{219,68,56}
\definecolor{c2}{RGB}{74,91,163}
\definecolor{c3}{RGB}{105,182,106}

\theoremstyle{definition}

\newcolumntype{d}[1]{D{.}{.}{#1}}  

\begin{document}

\title{Dynamical Decoupling via Continuous Unitary Design}

\author{Xiaodong Yang}
\affiliation{Institute of Quantum Precision Measurement, State Key Laboratory of Radio Frequency Heterogeneous Integration, College of Physics and Optoelectronic Engineering, Shenzhen University, Shenzhen 518060, China}
\affiliation{Quantum Science Center of Guangdong-Hong Kong-Macao Greater Bay Area (Guangdong), Shenzhen 518045, China}


\author{Jun Li}
\email{lijunquantum@szu.edu.cn}
\affiliation{Institute of Quantum Precision Measurement, State Key Laboratory of Radio Frequency Heterogeneous Integration, College of Physics and Optoelectronic Engineering, Shenzhen University, Shenzhen 518060, China}
\affiliation{Quantum Science Center of Guangdong-Hong Kong-Macao Greater Bay Area (Guangdong), Shenzhen 518045, China}

\begin{abstract}
Dynamical decoupling is a cornerstone of quantum control for protecting and engineering quantum systems. However, conventional schemes rely on idealized instantaneous pulses, limiting their performance under realistic finite-duration and imperfect controls. To overcome this limitation, we establish a dynamical decoupling framework based on continuous unitary designs, which intrinsically suppresses arbitrary static noise during continuous control. We further derive closed-form phase conditions to suppress Rabi-amplitude errors, unifying universal decoupling, finite-pulse compatibility, and control-error robustness within a single analytical framework. We experimentally validate our approach on a superconducting quantum cloud platform, demonstrating significantly extended coherence times and the highest robustness against Rabi-amplitude errors and detuning among the tested sequences.
\end{abstract}

\maketitle

\textit{Introduction}---Dynamical decoupling (DD) is a powerful tool for preserving quantum coherence and manipulating quantum dynamics \cite{SDA16}. By applying appropriately designed $\pi$-pulse sequences interleaved with free evolution, DD suppresses unwanted system-environment interactions, enabling applications in quantum memory \cite{Du09,wang2017single,PhysRevX.9.031045,ma2021one}, quantum sensing \cite{PhysRevX.5.041016,PhysRevX.8.021059,PhysRevLett.131.100801,PhysRevLett.131.220803,PhysRevLett.132.250801}, and noise-spectrum characterization \cite{PhysRevLett.107.230501,PhysRevLett.107.170504,PRXQuantum.2.010316,davis2023probing}. Over the past decades, numerous DD protocols have been developed to improve decoupling efficiency and robustness \cite{PhysRevLett.105.200402,PhysRevLett.104.130501,PhysRevLett.106.240501,GGT17,PhysRevLett.122.010408,WZL19,PhysRevApplied.18.054075,fk7j-y1vl}. More recently, DD has been extended beyond single-qubit coherence preservation to more general tasks, ranging from quantum state or gate protection in multilevel and multiqubit systems \cite{PhysRevLett.121.220502,PhysRevLett.123.010501,PhysRevApplied.18.024068,PhysRevLett.131.210802,6wd3-wc5s,PhysRevLett.134.050601} to many-body Hamiltonian engineering \cite{CJZ20,PhysRevX.14.031017}. These advances have established DD as a versatile tool for robust control in noisy intermediate-scale quantum  devices.

DD relies on effective averaging of unwanted interactions through structured control sequences, motivating a variety of averaging strategies. Early schemes, including the Hahn echo \cite{PhysRev.80.580}, Carr–Purcell \cite{PhysRev.94.630}, and Carr-Purcell-Meiboom-Gill \cite{meiboom1958modified} sequences, achieved deterministic averaging of dephasing through single-axis refocusing pulses. This principle was later formalized through group-theoretical constructions, with periodic \cite{PhysRevLett.82.2417} and concatenated \cite{PhysRevLett.95.180501} DD employing multi-axis pulses to achieve universal decoupling based on Pauli-group averaging, as exemplified by XY4. Subsequent phase-engineered sequences, such as Knill DD \cite{PhysRevLett.105.200402,PhysRevLett.106.240501} and universally robust (UR) DD \cite{GGT17}, improved the robustness of this averaging process. Other approaches optimized pulse timing and filter functions, as in Uhrig DD \cite{PhysRevLett.98.100504} and quadratic DD \cite{PhysRevLett.104.130501}, through analytical design or learning-based protocols on quantum hardware \cite{biercuk2009optimized,PhysRevX.8.021059}. More recently, randomized DD \cite{WZL19,fk7j-y1vl} introduced stochastic averaging over different sequence realizations to suppress coherent error accumulation. Despite these advances, conventional DD sequences remain within the bang-bang framework \cite{PhysRevLett.82.2417,PhysRevLett.95.180501,PhysRevLett.83.4888}, which assumes ideal instantaneous $\pi$ pulses. In practice, however, finite-duration pulses and experimental imperfections limit the performance of averaging strategies based on such idealized control operations \cite{PhysRevApplied.20.064027}.

In this work, we introduce continuous-unitary-design-inspired dynamical decoupling (CUDD), a fundamentally different averaging framework compatible with realistic finite-duration and imperfect control pulses. Continuous unitary designs (CUDs)~\cite{yang2026continuous} are continuous trajectories in the unitary group whose time averages reproduce the corresponding Haar averages, ensuring intrinsic first-order robustness against arbitrary static noise. Building on this property, we construct CUDD sequences by concatenating CUD blocks with free-evolution intervals at their quarter points, achieving universal decoupling without the instantaneous-pulse approximation. We further derive closed-form phase conditions to suppress Rabi-amplitude errors, providing simultaneous robustness against arbitrary static noise and systematic control imperfections. We experimentally benchmark CUDD against conventional DD sequences on a superconducting quantum cloud platform, demonstrating improved coherence preservation over a broad range of pulse intervals and enhanced robustness against both Rabi-amplitude errors and detuning.

\begin{figure}
\begin{center}
\includegraphics[width=0.99\linewidth]{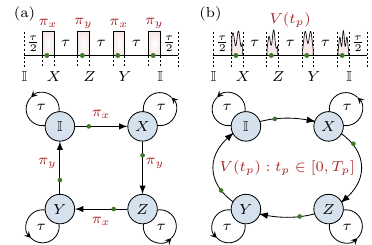}
\end{center}	
\caption{Comparison of XY4 and its CUDD counterparts. 
(a) Realistic XY4 uses alternating finite-width $\pi_x$ and $\pi_y$ pulses separated by free evolution $\tau$ ($\tau/2$ at ends). Ideal infinitely-short pulses permute Pauli group elements into a unitary 1-design, enabling complete decoupling via group symmetrization (Cayley graph below). However, realistic finite-width pulses form the 1-design only at the pulse end, with deviations at intermediate times (e.g., green sampling points) degrading decoupling performance.
(b) CUDD forms a CUD path $V(t_p)$ that intersects the Pauli group (up to a phase) at quarter-points of the total pulse time $T_p$ (bottom graph). Crucially, this path ensures that any set of four symmetric points (green dots) sampled during the pulse still forms a unitary 1-design, making CUDD intrinsically robust against arbitrary static noise throughout the pulses.
 }\label{example}
\end{figure}

\textit{Motivating Example}---We use the canonical XY4 sequence, shown in Fig.~\ref{example}(a), to illustrate its DD mechanism and practical limitations. Ideally, instantaneous $\pi$ pulses in XY4 traverse  the Pauli group $\mathcal{G}=\{\mathbb{I},X,Y,Z\}$, where $X\equiv \sigma_x , Y \equiv \sigma_y$ and $ Z\equiv \sigma_z$ denote the Pauli matrices and $\mathbb{I}$ represents the identity matrix. By uniformly covering the group elements, these ideal pulses form a unitary 1-design. This enables the group symmetrization to project an arbitrary noise operator  $\mathcal{N}$ onto the centralizer of $\mathcal{G}$ via the uniform average 
$\bar{\mathcal{N}}=\frac{1}{|\mathcal{G}|}\sum_{g_j\in\mathcal{G}}g_j^\dagger \mathcal{N} g_j$ \cite{PhysRevLett.82.2417,PhysRevLett.95.180501}, where $g_j$ are the group elements and $|\mathcal{G}|$ is the group order. For traceless noise, this yields exact first-order decoupling. This  mechanism is visualized as discrete permutations between the group elements; see Fig.~\ref{example}(a). However, realistic   pulses have finite widths, causing the evolution to traverse continuous paths in $\mathrm{SU}(2)$ rather than discrete group elements and violating the unitary 1-design condition at intermediate points (green dots), thereby introducing errors and degrading DD performance.

CUDD overcomes this limitation by replacing the four discrete $\pi$ pulses in XY4 with a  CUD path \cite{yang2026continuous} $V(t_p)=e^{-i\Omega t_p \sigma_z/2} e^{-i\Omega t_p \sigma_{3\pi/4}}$ over the total duration $T_p=2\pi/\Omega$, where $\Omega$ denotes the Rabi frequency and $\sigma_{\theta}\equiv\cos(\theta)\sigma_x+\sin(\theta)\sigma_y$. This path is divided into four equal-time segments, each of which replaces one original $\pi$ pulse. At the quarter-points (i.e., the end of each segment), the path yields $V(T_p/4)=i\sigma_x,  V(T_p/2)= i\sigma_z, V(3T_p/4)= -i\sigma_y, V(T_p)=-\mathbb{I}$, thereby constituting the Pauli group up to a global phase, as illustrated in Fig.~\ref{example}(b). 
More importantly, this continuous path preserves the unitary 1-design at every four sampled symmetry points, guaranteeing intrinsic first-order robustness to arbitrary static noise during finite-width pulses~\cite{yang2026continuous}. Hence, CUDD ensures effective decoupling under realistic finite-width pulses.

\textit{General CUDD framework}---Consider a quantum system $S$ coupled to an environment $E$, with Hamiltonian $H_0 = H_S \otimes \mathbb{I}_E + \mathbb{I}_S \otimes H_E + \sum_\alpha S_\alpha \otimes E_\alpha$. Here, $\mathbb{I}_{S,E}$ denote the identity operators  and $S_\alpha$ ($E_\alpha$) acts on $S$ ($E$), with $S_\alpha$ being traceless. 
DD aims to suppress system-environment interactions by applying a control Hamiltonian $H_c(t)\otimes \mathbb{I}_E$ that acts only on $S$. The corresponding control propagator is $U_c(t)=\mathcal{T}\exp\{-i\int_0^t H_c(t')dt'\}$. In the toggling frame, the total evolution operator becomes $U(t)=U_c(t)\mathcal{T}\exp\left\{-i\int_0^t \tilde H_0(t')dt'\right\}$, where $\tilde H_0(t)=U_c^\dag(t)H_0U_c(t)$. 
For a cyclic control sequence of period $T$ satisfying $U_c(T)=\mathbb{I}$, average-Hamiltonian theory gives $ U(T)=\exp(-i\overline{H}T)$, where $\overline{H}=\overline{H}^{(0)}+\overline{H}^{(1)}+\cdots$ and
\begin{equation}\label{h0}
\overline{H}^{(0)}=\frac{1}{T}\int_0^{T}U_c^\dag(t)H_0U_c(t)dt.
\end{equation}
Conventional DD treats pulses as instantaneous, whereas realistic finite-width pulses induce continuous control evolution that must be accounted for.

\begin{figure}
\begin{center}
\includegraphics[width=0.9\linewidth]{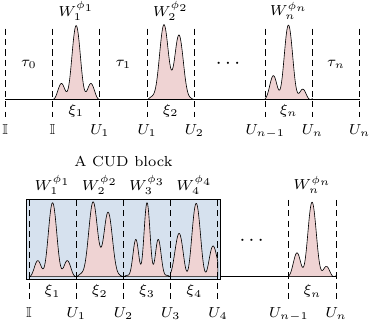}
\end{center}	
\caption{Schematic illustration of the CUDD framework, consisting of $n$ finite-duration pulses of durations $\xi_k$, separated by free-evolution intervals $\tau_k$, where $k=1,2,...,n$. The framework combines universal decoupling with intrinsic robustness against arbitrary static noise by ensuring that every four cumulative pulse evolutions form a CUD path. Further robustness against Rabi-amplitude errors is achieved by introducing an independent phase shift $\phi_k$ for each pulse, chosen to satisfy the corresponding analytical conditions.}\label{scheme}
\end{figure}

To incorporate this continuous evolution, we consider a sequence of $n$ finite-duration  pulses interspersed with free-evolution intervals $\tau_k$; see Fig.~\ref{scheme}.  The $k$th pulse is generated by a control field $\boldsymbol{u}_k(s)$ of duration $\xi_k$, with pulse propagator $W_k(s)=\mathcal{T}\exp(-i\int_0^s H_c(\boldsymbol{u}_k(s'))ds')$ and $W_k\equiv W_k(\xi_k)$. The total sequence duration is $T=\sum_{k=0}^n\tau_k+\sum_{k=1}^n\xi_k$. Substituting this  sequence into Eq.~(\ref{h0}) yields the modified zeroth-order effective Hamiltonian (see Supplemental Material)
\begin{equation}\label{eff}
H_{\mathrm{eff}} = \frac{1}{T}\left( \sum_{k=0}^{n} \tau_k U_k^\dagger H_0 U_k + \int_0^{T_p} dt_p  U_p^\dagger(t_p) H_0 U_p(t_p) \right),
\end{equation}
where $U_k=W_k\cdots W_1$ ($U_0=\mathbb{I}$) is the cumulative evolution after $k$ complete pulses, and $U_p(t_p)$ is the cumulative pulse evolution parametrized by the total pulse time $t_p\in[0,T_p]$ ($T_p=\sum_k\xi_k$). 
Equation (\ref{eff})  highlights two contributions to realistic DD: the discrete group-theoretic averaging determined by the propagators at the pulse boundaries, and the continuous averaging accumulated during the finite-duration pulses. Effective decoupling therefore requires both the appropriate boundary propagators and a suitably designed continuous control trajectory during each pulse.

To simultaneously satisfy both requirements, we group every four pulses to form a CUD path
\cite{yang2026continuous} (see Fig.~\ref{scheme})
\begin{equation}\label{cud_path}
V(t_p)=e^{-i\Omega t_p \sigma_z/2} e^{-i\Omega t_p \sigma_{\phi_0}},
\end{equation}
where $\phi_0$ denotes an arbitrary initial phase. We further assume uniform pulse durations $\xi_k=\xi=\pi/(2\Omega)$ and intervals $\tau_k=\tau$ (with boundary intervals $\tau_0=\tau_n=\tau/2$), and set $\phi_0=\pi/2$. At the four quarter-points of the block, the cumulative evolution yields  $V(\xi) = (-i\sigma_y - \sigma_x)/{\sqrt{2}}, V(2\xi) = i\sigma_z, V(3\xi) = (-i\sigma_y + \sigma_x)/{\sqrt{2}}, V(4\xi) = -\mathbb{I}$. Crucially, these boundary propagators also form a unitary 1-design over $\mathrm{SU}(2)$. Consequently, the discrete average over the four quarter-points vanishes exactly for any traceless $H_0$, i.e, $\sum_{k=1}^4 V^\dagger(k\xi) H_0 V(k\xi) = 0$. Repeating this structure for all $M$ CUD blocks,  the first term in Eq.~\eqref{eff} is thus eliminated across the $n=4M$ pulses.

On the other hand, this CUD path satisfies the condition  \cite{yang2026continuous}  
$
\frac{1}{4\xi}\int_0^{4\xi} V^\dagger(t_p) \mathcal{N}  V(t_p) dt_p
= \int_{\mathrm{SU}(2)} U^\dagger \mathcal{N}  U dU
= \frac{\mathbb{I}}{2}\operatorname{Tr}(\mathcal{N} )
$
for any operator $\mathcal{N} $ , where $dU$ denotes the normalized Haar measure. 
Since the environmental coupling in $H_0$ is traceless,  setting $\mathcal{N} = H_0$ in the above equation guarantees that the full pulse integral over the $M$ CUD blocks is zero ($T_p=4M\xi$):
$
\int_0^{T_p}U_p^\dagger(t_p) H_0 U_p(t_p) dt_p = M \int_{0}^{4\xi}V^\dagger(t_p) H_0 V(t_p) dt_p = 0
$.
Consequently, the second term in Eq.~(\ref{eff}) vanishes exactly. 
Overall, we refer to this sequence as CUDD, and use CUDD$n$ to denote a CUDD sequence comprising $n$ pulses.

Physically, the $k$th pulse is driven by the control Hamiltonian $H_k(s) = \sum_{\mu} u_{k,\mu}(s) \sigma_\mu / 2$ ($\mu=x,y,z; k=1,...,4M$), where $s \in [0, \xi]$ is the segment-local time. Solving $\dot{V}(t_p)=-iH_k(t_p)V(t_p)$ on $t_p \in [(j-1)\xi, j\xi]$ yields the piecewise control fields for the $k$th pulse:
\begin{equation}\label{cudd_pulse}
\boldsymbol{u}_k(s)=\left(2\Omega\cos(\Omega s+\varphi_k), 2\Omega\sin(\Omega s+\varphi_k),\Omega\right),
\end{equation}
where $j = ((k-1) \bmod 4) + 1$ denotes the in-block position and $\varphi_k =  \pi/2 + \frac{\pi}{2}((k-1) \bmod 4)$.
It is worth noting that CUDD is distinct from Eulerian DD \cite{PhysRevLett.90.037901,PhysRevLett.102.080501}, which employs bounded-strength controls arranged along Eulerian cycles, at the cost of substantially increased control complexity.

\textit{Robust CUDD}---In practice, the time-dependent transverse control fields are susceptible to Rabi errors. To compensate for these errors, we introduce an independent phase shift $\phi_k$ for each pulse, yielding the modified control fields 
\begin{equation}
\boldsymbol{u}_k^{\phi_k}(s) = (  2\Omega\cos(\Omega s+\Phi_k), 
 2\Omega\sin(\Omega s+\Phi_k), \Omega),
\end{equation}
where $\Phi_k=\varphi_k+\phi_k$. Thus, in the rotating frame defined by $e^{-i\Omega t\sigma_z/2}$, the phase-corrected control Hamiltonian with Rabi errors takes the form $ H_k^{\phi_k}=(1+\varepsilon)\Omega[\cos(\Phi_k)\sigma_x+\sin(\Phi_k)\sigma_y]=(1+\varepsilon)\Omega \sigma_{\Phi_k}$, where $\varepsilon$ is the strength of the Rabi error. The corresponding evolution operator over the $k$th segment of duration $\xi$ is $ W_k^{\phi_k} = e^{-i \xi (1+\varepsilon)\Omega \sigma_{\Phi_k}}$, and the overall sequence propagator is $U_{4M}=\prod_{k=1}^{4M} W_k^{\phi_k}$.

To retain the universal decoupling and intrinsic robustness against arbitrary static noise of CUDD, the phase-corrected pulse evolution must preserve the continuous unitary 1-design property: $\int_0^{T_p} dt_p [U_p^{\phi_k}(t_p)]^\dagger \sigma_u U_p^{\phi_k}(t_p) =\sum_{k=1}^{4M}\int_{(j-1)\xi}^{j\xi} [ V_k^{\phi_k}(t_p)]^\dag \sigma_u  V_k^{\phi_k}(t_p) d t_p =0, \mu=x,y,z, j = ((k-1) \bmod 4) + 1$, where $U_p^{\phi_k}(t_p)$ and $V_k^{\phi_k}(t)$ denote the phase-corrected cumulative pulse evolution and CUD path, respectively. In addition, the sequence fidelity $F=\frac{1}{4}\big|\mathrm{Tr}\big(U_{\mathrm{target}}^{\dagger}U_{4M}\big)\big|^{2}$ should be maximized to ensure robustness against Rabi-amplitude errors, where $U_{\mathrm{target}}=\pm\mathbb{I}$. Furthermore, we apply symmetric block ansatz: within each block~$m$, the four phases share a common base $\alpha_m$ and differ only by a fixed offset pattern $\boldsymbol\delta=(\delta_1,\delta_2,\delta_3,\delta_4)$, namely $\phi_k=\phi_{4(m-1)+j}=\alpha_m+\delta_j$, where $m=1,...,M$.

\begin{table}
\centering
\renewcommand{\arraystretch}{1.5}
\begin{tabular}{lll}
\hline\hline
Sequence & Phases $\{\phi_k\}$ & $\{\alpha_m\}$ \\
\hline
CUDD8  & $\pi(0,1,0,1,\;1,0,1,0)$
       & $\pi\{0,1\}$ \\

CUDD12 & $\pi(0,1,0,1,\;\frac{2}{3},\frac{5}{3},\frac{2}{3},\frac{5}{3},\;\frac{4}{3},\frac{1}{3},\frac{4}{3},\frac{1}{3})$
       & $\frac{2\pi}{3}\{0,1,2\}$ \\

CUDD16 & $\frac{\pi}{2}(0,2,0,2,\;1,3,1,3,\;2,0,2,0,\;3,1,3,1)$
       & $\frac{\pi}{2}\{0,1,2,3\}$ \\

\hline\hline
\end{tabular}
\caption{Representative robust CUDD sequences with $M=2,3$ and $4$ blocks. The phases are constructed using uniformly spaced base phases $\alpha_m=(m-1)2\pi/M$ and the offset pattern $\boldsymbol{\delta}=(0,\pi,0,\pi)$, such that $\phi_k=\phi_{4(m-1)+j}=\alpha_m+\delta_j$. All phases are defined modulo $2\pi$.}
\label{table1}
\end{table}

The above requirements lead to the following analytical conditions for robust CUDD sequences:
\begin{equation}
	\sum_{m=1}^{M}e^{i\alpha_m}=0,  \sum_{j=1}^{4}e^{i(\tilde \delta _j+\tilde \delta _{j-1})}=0, \sum_{j=1}^4 (-1)^{j-1}\delta_j = j'\pi,
\end{equation}
where $j' \in\mathbb{Z}$ and $ \tilde \delta _j=\sum_{l=1}^j(-1)^{l-1}\delta_l, \tilde \delta _0=0 $.
The first requires that the $M$ unit vectors in the complex plane form a closed polygon, whereas the latter two are automatically satisfied by any offset pattern $\boldsymbol{\delta}$ containing exactly two zeros and two $\pi$'s. Specifically, explicit phase solutions for representative CUDD sequences are listed in Table~\ref{table1}, with details and additional examples provided in Supplemental Material.

\begin{figure*}
\begin{center}
\includegraphics[width=0.98\linewidth]{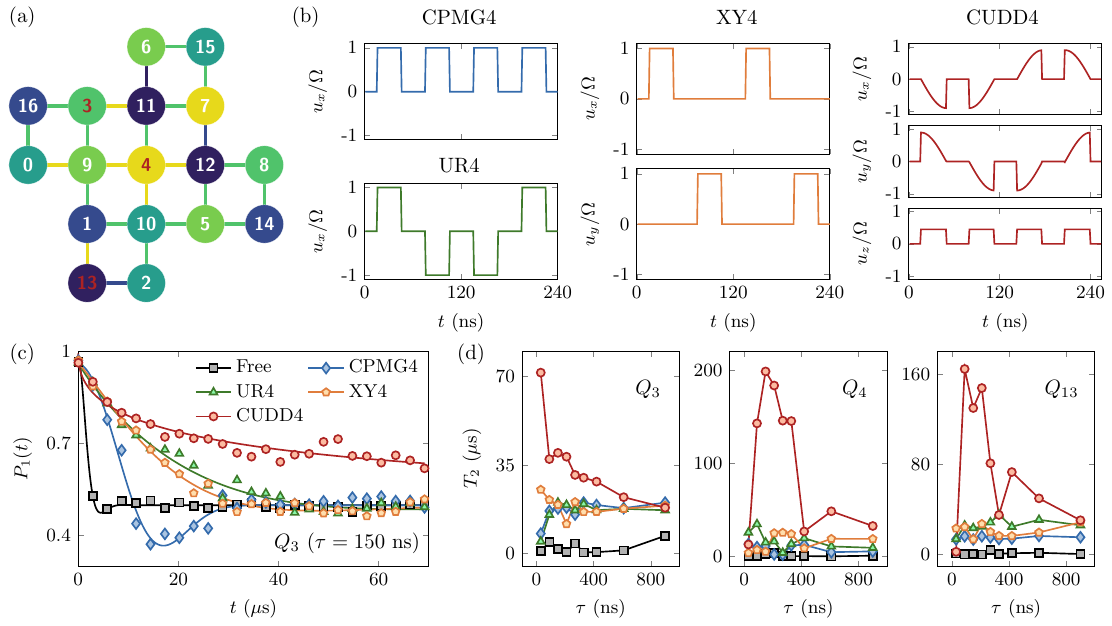}
\end{center}	
\caption{Experimental comparison of CUDD with conventional DD sequences. (a) Connectivity graph of the superconducting chip QZ01. The red-marked qubits $Q_3$, $Q_4$, and $Q_{13}$ were the ones tested in the experiments. (b) Waveforms of the different DD sequences with $\xi = 30$ ns and $\Omega$ representing the Rabi frequency. The case of $\tau = 30$ ns is explicitly shown for illustration. For each DD sequence, the control fields along the $x$, $y$, and $z$ axes are denoted by $u_x$, $u_y$, and $u_z$, respectively, with any unplotted component being zero. (c) Probability of measuring state $|1\rangle$ as a function of $t$ under different sequences. The markers represent experimental data, and the solid lines are fits to $S(t) = C_{\text{inf}} + A \cos(\omega t + \Delta) \exp\left[ -(t/T_2)^p \right]$. (d) The fitted coherence time $T_2$ as a function of the pulse interval $\tau$ for different sequences tested on qubits $Q_3$, $Q_4$, and $Q_{13}$. The markers are experimental results, and the solid lines simply connect these markers for visual guidance. }\label{dd_compare}
\end{figure*}

\textit{Experiments.}---We first experimentally benchmark the CUDD sequence on the superconducting quantum cloud developed by \textit{Logical Qubit} \cite{logical_qubit_cloud}. The experiments are performed on the 17-qubit QZ01 chip, whose connectivity graph is shown in Fig.~\ref{dd_compare}(a). We select qubits $Q_3$, $Q_4$, and $Q_{13}$ for benchmarking because they have different connectivity patterns with their neighboring qubits, allowing us to assess the performance of CUDD under different qubit-connectivity conditions. We followed standard procedures to calibrate the chip parameters; details are provided in the Supplemental Material. In the experiments, we set $\xi=30$ ns.

For comparison, we benchmark several commonly used DD sequences, whose waveforms are shown in Fig.~\ref{dd_compare}(b). The control fields along the $x,y$ and $z$ axes are denoted by $u_x,u_y$ and $u_z$, respectively. Any control-field component not shown in the plot is zero. In addition, we rescale the CUDD control fields to ensure that their maximum amplitude does not exceed the Rabi frequency. Accordingly, the control pulses in Eq.~(\ref{cudd_pulse}) become $\boldsymbol{u}_k(s)=\Omega/\sqrt{5} \cdot \left(2\cos(\Omega s/\sqrt{5}+\varphi_k),\,2\sin(\Omega s/\sqrt{5}+\varphi_k),\,1\right)$, where $\varphi_k=\pi/2+\frac{\pi}{2}[(k-1)\bmod 4]$ and $s\in[0,2\pi \sqrt{5}/\Omega ]$. This rescaling makes CUDD sequence slightly longer in duration. 

The benchmarking procedure is as follows. The tested qubit is initialized in $|0\rangle$, rotated by $\pi/2$ about the $x$ axis, and subjected to repeated applications of the DD sequence with fixed pulse interval $\tau$. A final $\pi/2$ rotation about the same axis is applied before computational-basis measurement. The probability of obtaining $|1\rangle$, $P_1(t)$, is recorded as a function of the total duration of the repeated DD sequence.
A representative result obtained for qubit $Q_3$ with $\tau=150$ ns is shown in Fig.~\ref{dd_compare}(c). The markers represent the experimental data, while the solid lines show fits to $S(t)=C_{\text{inf}}+A\cos(\omega t+\Delta)\exp[-(t/T_2)^p]$, where $T_2$ denotes the fitted coherence time (see fitted parameters  in Supplemental Material). The results clearly show that CUDD4 preserves coherence significantly better than the other DD sequences. 
Furthermore, we vary the pulse interval $\tau$ and measure the corresponding coherence time $T_2$ on qubits $Q_3$, $Q_4$, and $Q_{13}$; see Fig.~\ref{dd_compare}(d). It is evident that CUDD consistently achieves longer coherence times than the other DD sequences over most of the tested pulse intervals $\tau$ across all three qubits. The improvement is particularly pronounced at shorter pulse intervals, where errors accumulated during finite-duration pulses become more significant. In this regime, CUDD offers enhanced robustness against arbitrary static noise during the control pulses, leading to better coherence preservation. However, when $\tau$ becomes too short, as in the case of $\tau=30$ ns for $Q_4$ and $Q_{13}$, the performance of CUDD deteriorates and falls below that of some other sequences. This can be attributed to the increased pulse errors at such short intervals, which exceed the range over which the first-order robustness of CUDD remains effective. Conversely, at sufficiently large pulse intervals, errors arising from finite-duration pulses become less significant, and all DD sequences exhibit comparable performance.

\begin{figure}
\begin{center}
\includegraphics[width=0.98\linewidth]{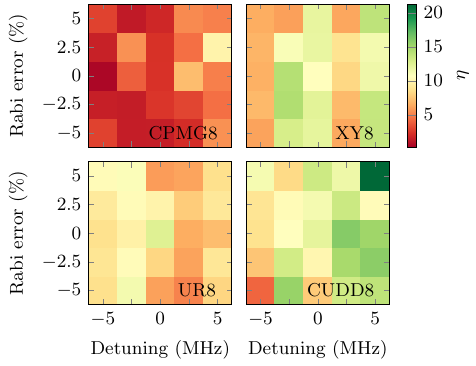}
\end{center}	
\caption{Experimental comparison of the robust CUDD sequence and conventional DD sequences in the presence of Rabi errors and detuning on $Q_4$, with $\tau=60$ ns. 
The coherence protection efficiency $\eta$ is defined as the ratio of the integrated coherent signal of a given DD sequence to that of the free evolution, with both signals measured relative to the 0.5 baseline. }\label{dd_robust}
\end{figure}

We next test the robustness of CUDD8 against Rabi errors and detuning on $Q_4$ with $\tau=60$ ns. Since the intrinsic Rabi error of the chip is negligible, we deliberately introduce Rabi errors and compare CUDD8 with CPMG8, XY8, and UR8. Given its intrinsic robustness against arbitrary static noise during the pulses, CUDD is also expected to be robust against detuning. The experimental results are shown in Fig.~\ref{dd_robust}. We define the coherence protection efficiency as $\eta=|\int_0^T[P_1(t)-0.5]dt|/|\int_0^T[P_{\text{free}}(t)-0.5]dt|$, where $T$ is the total sequence time for 40 repetitions and  $P_{\text{free}}(t)$ denotes the measured probability in the absence of DD sequences. We find that XY8 is more robust than CPMG8 because it averages errors along all three axes, whereas CPMG8 addresses only two. UR8 performs slightly worse than XY8, which can be attributed to the relatively short pulse interval $\tau$, consistent with the observation in Ref.~\cite{GGT17}. Overall, CUDD8 generally exhibits the highest robustness, as it is specifically designed to suppress both types of pulse imperfections.

\textit{Conclusion and outlook.}---In this work, we introduce CUDD as a continuous-averaging framework for DD that goes beyond the conventional bang-bang paradigm. By exploiting the averaging properties of CUDs, we achieve universal decoupling with finite-duration pulses and further enhance its robustness against Rabi-amplitude errors through tailored phase corrections. These results are experimentally validated on a superconducting quantum cloud platform.

The proposed CUDD sequence is general and experimentally accessible: its pulse envelopes have simple sinusoidal or cosinusoidal forms, making them smooth and straightforward to implement on a wide range of quantum platforms. Future work could extend CUDD to multiqubit systems by incorporating graph coloring \cite{8q24-zj73} and Hadamard-matrix techniques \cite{1d4l-73x6}, enabling scalable protection of coherence and entanglement against complex unwanted interactions. In addition, CUDD can be directly applied to other DD-based applications, such as quantum sensing and noise-spectrum characterization. Its robustness against pulse errors is particularly beneficial for high-frequency signal detection, where short pulse intervals amplify their impact.

\emph{Acknowledgments---}\noindent
We gratefully acknowledge \textit{Logical Qubit}  for providing quantum cloud platform services. We acknowledge support by Quantum Science and Technology-National Science and Technology Major Project (Grant No. 2024ZD0300400), the Guangdong Provincial Quantum Science Strategic Initiative (Grant No. GDZX2505001), the National Natural Science Foundation of China (Grants No. 12675023, 12404554 and 12441502), Guangdong Basic and Applied Basic Research Foundation (Grant No. 2026A1515010331),  the Shenzhen Science and Technology Program (Grant No. ZDYJ20251211120900001).

\emph{Data availability---}\noindent
The datasets generated during and/or analysed during the current study are available from the corresponding author on reasonable request.


%


\clearpage
\onecolumngrid

\setcounter{equation}{0}
\setcounter{figure}{0}
\setcounter{table}{0}

\section*{Supplementary Material}


%
%

\section{Derivation of the Effective Hamiltonian Expression}\label{efffective_ham}
Consider an $n$-pulse decoupling sequence with total duration $T$, consisting of $n$ control pulses of durations $\xi_k$ ($k=1,\ldots,n$) and $n+1$ free-evolution intervals of durations $\tau_k$ ($k=0,\ldots,n$), such that 
\begin{equation}
	T=\sum_{k=0}^n \tau_k +\sum_{k=1}^n \xi_k.
\end{equation}
For convenience, we define the starting time of the $k$th pulse as $t_k=\sum_{i=0}^{k-1}\tau_i+\sum_{i=1}^{k}\xi_i -\xi_k$ with $t_0=\xi_0=0$.

To describe the dynamics induced by the control field during the $k$th pulse, we introduce a local-time evolution operator as
\begin{equation}
	W_k(s) = \mathcal{T} \exp\left( -i\int_0^{s} ds' H_C(\boldsymbol{u_k}(s')) \right), \quad s \in [0,\xi_k].
\end{equation}
Meanwhile, the corresponding global-time evolution can be written as
\begin{equation}
	W_k^{(g)}(t) = \mathcal{T} \exp\left( -i\int_{t_k}^{t} dt' H_C(\boldsymbol{u_k}(t')) \right), \quad t \in [t_k,t_k + \xi_k].
\end{equation}

To characterize the overall dynamics generated by this sequence, we employ the  average Hamiltonian theory. According to the main text, the corresponding first-order effective Hamiltonian is given by
\begin{equation}
H_{\mathrm{eff}}= \frac{1}{T}\int_0^T dt  U_c^\dagger(t) H_0 U_c(t),
\end{equation}
where $H_0$ is the joint Hamiltonian of the system and the environment, and $U_c(t)$ is the control evolution operator. We further define the following evolution operators
\begin{itemize}
\item $W_k$: evolution operator for the $k$th pulse, $W_k \equiv W_k(\xi_k)$
\item $U_k \equiv W_k W_{k-1} \cdots W_1$: cumulative evolution after $k$ pulses ( $U_0 = \mathbb{I}$ and $\mathbb{I}$ denotes the identity matrix)
\item $U_k(s) \equiv W_k(s) U_{k-1}$: cumulative evolution in local time during the $k$th pulse,  $s \in [0, \xi_k]$
\item $U_k^{(g)}(t) \equiv W_k^{(g)}(t) U_{k-1}$:  cumulative evolution in global time during the $k$th pulse,  $t \in [t_k, t_k+\xi_k]$
\end{itemize}

\subsection{Decomposition of the integral}
During the $k$th pulse interval $I_k^{\mathrm{pulse}}=[t_k, t_k + \xi_k]$, the evolution operator is time-dependent
\begin{equation}
 U_c(t)=U_k^{(g)}(t) = W_k^{(g)}(t) U_{k-1}=W_k(t-t_k) U_{k-1}, \quad t \in [t_k, t_k + \xi_k].
\end{equation}
The contribution from the $k$th pulse is
\begin{align}
	\int_{t_k}^{t_k + \xi_k} dt U_c^\dagger(t) H_0 U_c(t) &= \int_{t_k}^{t_k + \xi_k} dt  U_{k-1}^\dagger W_k^\dagger(t-t_k) H_0 W_k(t-t_k) U_{k-1} \nonumber\\ &= \int_0^{\xi_k} ds U_{k-1}^\dagger W_k^\dagger(s) H_0 W_k(s) U_{k-1} \nonumber\\ 
	&=\int_0^{\xi_k} ds U_k^\dagger(s) H_0 U_k(s),
\end{align}
where we used the substitution $s = t - t_k$. Summing over all pulse intervals gives
\begin{equation}
\sum_{k=1}^{n} \int_0^{\xi_k} ds U_k^\dagger(s) H_0 U_k(s), \quad s \in [0,\xi_k].
\end{equation}

During the $k$th free-evolution interval $I_k^{\mathrm{free}}=[t_k+\xi_k,\,t_k+\xi_k+\tau_k]$ of duration $\tau_k$, the evolution operator remains constant at $U_k$. The contribution is
\begin{equation}
\int_{I_k^{\text{free}}} dt\, U_c^\dagger(t) H_0 U_c(t) = \tau_k U_k^\dagger H_0 U_k.
\end{equation}
Summing over all free-evolution intervals gives
\begin{equation}
\sum_{k=0}^{n} \tau_k U_k^\dagger H_0 U_k.
\end{equation}

Therefore, the complete decomposition is
\begin{equation}
\int_0^T dt  U_c^\dagger(t) H_0 U_c(t) = \sum_{k=0}^{n} \tau_k U_k^\dagger H_0 U_k + \sum_{k=1}^{n} \int_0^{\xi_k} ds  U_k^\dagger(s) H_0 U_k(s).
\end{equation}

\subsection{Continuous integral representation}
To rewrite the pulse contributions as a single continuous integral, we introduce a cumulative pulse-time variable $t_p \in [0, T_p]$, where $T_p = \sum_{k=1}^n \xi_k$ denotes the total duration of all pulses. This allows us to recast the sum of pulse integrals as
\begin{align}\label{eq11}
	\sum_{k=1}^{n} \int_0^{\xi_k} ds  U_k^\dagger(s) H_0 U_k(s)
	& = \sum_{k=1}^{n} \int_0^{\xi_k} ds  U_{k-1}^\dagger W_k^\dagger(s) H_0 W_k(s) U_{k-1} \nonumber\\
	&=  \sum_{k=1}^{n} \int_{\sum_{i=1}^{k-1} \xi_i}^{\sum_{i=1}^{k}\xi_i} dt_p  U_{k-1}^\dagger W_k^\dagger(t_p-\sum_{i=1}^{k-1}\xi_i) H_0 W_k(t_p-\sum_{i=1}^{k-1}\xi_i) U_{k-1} \nonumber\\
	&=  \sum_{k=1}^{n} \int_{\sum_{i=1}^{k-1} \xi_i}^{\sum_{i=1}^{k}\xi_i} dt_p  U_{k}^\dagger (t_p-\sum_{i=1}^{k-1}\xi_i) H_0 U_k(t_p-\sum_{i=1}^{k-1}\xi_i)  \nonumber\\
	&\equiv \int_0^{T_p} dt_p U_p^\dag(t_p)H_0  U_p(t_p),
\end{align}
where $ U_p(t_p)$ is defined piecewise for $t_p \in [0, T_p]$, i.e.,
\begin{equation}
 U_p(t_p) = 
\begin{cases}
U_1(t_p) = W_1(t_p), & 0 \leq t_p < \xi_1 \\
U_2(t_p - \xi_1) = W_2(t_p - \xi_1) U_1, & \xi_1 \leq t_p < \xi_1 + \xi_2 \\
\vdots \\
U_k(t_p - \sum_{i=1}^{k-1}\xi_i) = W_k(t_p - \sum_{i=1}^{k-1}\xi_i) U_{k-1}, & \sum_{i=1}^{k-1}\xi_i \leq t_p < \sum_{i=1}^{k}\xi_i
\end{cases}
\end{equation}

\subsection{Final result}
Combining both contributions, the effective Hamiltonian is
\begin{equation}
H_{\mathrm{eff}} = \frac{1}{T}\left( \sum_{k=0}^{n} \tau_k U_k^\dagger H_0 U_k + \int_0^{T_p} dt_p U_p^\dagger(t_p) H_0 U_p(t_p) \right),
\end{equation}
where $U_p(t_p) = U_k(t_p - \sum_{i=1}^{k-1}\xi_i)= W_k(t_p - \sum_{i=1}^{k-1}\xi_i) U_{k-1}$ for $t_p\in [\sum_{i=1}^{k-1}\xi_i, \sum_{i=1}^{k}\xi_i]$ (the $k$th pulse interval).

\section{Construction of the robust CUDD Sequences}
As described in the main text, to construct the CUDD sequence, we connect $M$ ($M\in \mathbb{N}^+$) continuous unitary design (CUD) blocks, each divided into four segments of equal duration $\xi = \pi/(2\Omega)$. For the $k$th segment ($k=1,2,...,4M$), where the global index is $k=4(m-1)+j$ (with block index $m=1,2,\dots,M$ and in-block position $j=1,2,3,4$), the cumulative pulse evolution over the block is a continuous path $V(t_p)$ at block-local time $t_p \in [0, 4\xi]$, given by
\begin{equation}
	V(t_p)=e^{-i \Omega t_p \sigma_z/2} e^{-i \Omega t_p \sigma_{\phi_0}}, 
\end{equation}
where $\sigma_{\phi_0}\equiv \sigma_x \cos\phi_0+ \sigma_y \sin\phi_0$ and $\sigma_x, \sigma_y, \sigma_z$ are the Pauli matrices.
Physically, the $k$th pulse is implemented by applying a control Hamiltonian $H_k(s) = \sum_{\mu} u_{k,\mu}(s) \sigma_\mu / 2 $ $(\mu=x,y,z)$, where $s \in [0, \xi]$ is the local time within the segment. Solving $\dot{V}(t_p)=-iH_k(t_p)V(t_p)$ restricted to the interval $t_p \in [(j-1)\xi, j\xi]$ yields the piecewise control fields for the $k$th pulse:
\begin{equation*}
	u_{k,x}(s)=2\Omega\cos[\Omega(s + (j-1)\xi)+\phi_0], \quad u_{k,y}(s)=2\Omega\sin[\Omega(s + (j-1)\xi)+\phi_0], \quad u_{k,z}(s)=\Omega.
\end{equation*}
Using $\Omega\xi = \pi/2$, we define the accumulated phase for the $k$th pulse as $\varphi_k \equiv  \phi_0 + (j-1)\frac{\pi}{2}=  \phi_0 + \frac{\pi}{2}((k-1) \bmod 4)$. The control fields simplify to:
\begin{equation*}
	u_{k,x}(s)=2\Omega\cos(\Omega s+\varphi_k), \quad u_{k,y}(s)=2\Omega\sin(\Omega s+\varphi_k).
\end{equation*}

The corresponding evolution operator for the $k$th pulse, measured from its starting time $t_p=(j-1)\xi$, is therefore given by $W_k(s)=V(t_p+s)V^\dagger(t_p)$, where $s\in[0,\xi]$ is the local time. By substituting the cumulative evolution path $V(t_p)$, we obtain:
\begin{equation}
W_k(s) = \left[ e^{-i\Omega (t_p+s) \sigma_z/2} e^{-i\Omega (t_p+s) \sigma_{\phi_0}} \right] \left[ e^{i\Omega t_p \sigma_{\phi_0}} e^{i\Omega t_p \sigma_z/2} \right].
\end{equation}
To analyze the local pulse, we shift to a frame rotating at the drive frequency $\Omega$. The effective evolution operator in this frame is defined by $U^\dagger(t_p+s) W_k(s) U(t_p)$ with $U(t) = e^{-i\Omega t \sigma_z/2}$. This yields:
\begin{equation}
{W}'_k(s) = e^{i\Omega (t_p+s) \sigma_z/2} W_k(s) e^{-i\Omega t_p \sigma_z/2} = e^{-i\Omega s \sigma_{\phi_0}}.
\end{equation}
This result indicates that the local rotation is referenced to the initial phase $\phi_0$. However, as derived from the physical control fields, the drive phase at the start of the $k$th segment has accumulated to $\varphi_k \equiv \phi_0 + \Omega t_p = \phi_0 + \frac{\pi}{2}((k-1) \bmod 4)$. By substituting this physical accumulated phase into the local rotation axis, the evolution operator for the $k$th pulse should take the form
\begin{equation}
{W}_k(s) = e^{-i\Omega s \sigma_{\varphi_k}}.
\end{equation}

In practice, the time-dependent transverse control fields are susceptible to Rabi errors. To compensate for this effect, we introduce a relative phase shift $\phi_k$ for each pulse. The transverse control fields become 
\begin{equation*}
	u_{k,x}^{\phi_k}(s)=2\Omega\cos(\Omega s+{\Phi}_k), \quad
	u_{k,y}^{\phi_k}(s)=2\Omega\sin(\Omega s+{\Phi}_k),
\end{equation*}
with $u_{k,z}^{\phi_k}(s)=\Omega$, where ${\Phi}_k = \varphi_k + \phi_k$. In the rotating frame defined by $e^{-i\Omega s\sigma_z/2}$, the phase-corrected control Hamiltonian with Rabi error is 
\begin{equation*}
	H_k^{\phi_k}=(1+\varepsilon)\Omega[\cos({\Phi}_k)\sigma_x+\sin({\Phi}_k)\sigma_y]=(1+\varepsilon)\Omega \sigma_{{\Phi}_k},
\end{equation*}
where $\varepsilon$ is the strength of the Rabi error.
As such, the evolution operator from the beginning to the end of segment $k$ (duration $\xi$), is
\begin{equation}
W_k^{\phi_k} = e^{-i \xi (1+\varepsilon)\Omega \sigma_{{\Phi}_k}}
= \begin{bmatrix}
\sqrt{1-p} & -i\sqrt{p}\,e^{-i{\Phi}_k} \\[4pt]
-i\sqrt{p}\,e^{i{\Phi}_k} & \sqrt{1-p}
\end{bmatrix},
\end{equation}
where $p = \sin^2[(1+\varepsilon)\pi/2]$ is the single-segment transition probability.

\subsection{Conditions for preserving the continuous unitary 1-design property}
The phase-corrected CUDD sequence should preserve  the continuous unitary 1-design property (see Eq. (\ref{eq11})), i.e., 
\begin{equation*}
	\int_0^{T_p} dt_p [U_p^{\phi_k}(t_p)]^\dagger \sigma_\mu U_p^{\phi_k}(t_p) =\sum_{k=1}^{4M}\int_0^\xi [ U_k^{\phi_k}(s)]^\dag \sigma_\mu  U_k^{\phi_k}(s) ds =0, \mu=x,y,z.
\end{equation*}
 Noticing that every four pulses in CUDD forms a CUD path, the above condition can be rewritten as 
 \begin{equation}\label{cu1d}
	I_\mu \equiv \sum_{k=1}^{4M} \int_{(j-1)\xi}^{j\xi} [V^{\phi_k}(t_p)]^\dagger \sigma_\mu V^{\phi_k}(t_p) dt_p = 0, \quad \mu = x, y, z,
\end{equation}
where  $V^{\phi_k}(t_p)\equiv R_z^\dag(\phi_k) V(t_p) R_z (\phi_k)$, $t_p \in [(j-1)\xi,j\xi]$ and $j = ((k-1) \bmod 4) + 1$. 
To proceed, the integrand in  the $k$th segment can be expressed as 
\begin{equation*}
	[V^{\phi_k}(t_p)]^\dagger \sigma_\mu V^{\phi_k}(t_p)= R_z^\dag(\phi_k)  V^\dag(t_p) R_z (\phi_k) \sigma_\mu R_z^\dag(\phi_k) V(t_p) R_z (\phi_k)\equiv R_z^\dag(\phi_k)  [V^\dag(t_p) \tilde \sigma_\mu(\phi_k) V(t_p)]R_z (\phi_k),
\end{equation*}
where $\tilde \sigma_\mu(\phi_k)=R_z (\phi_k) \sigma_\mu R_z^\dag(\phi_k)$, and explicitly $\tilde \sigma_x(\phi_k)=\sigma_x\cos \phi_k-\sigma_y \sin \phi_k, \tilde \sigma_y(\phi_k)=\sigma_x\sin \phi_k+\sigma_y \cos\phi_k, \tilde \sigma_z(\phi_k)=\sigma_z $.
Since the same position $j$ in different blocks corresponds to the same  evolution $V(t_p)$, we define 
\begin{equation}\label{eqabc}
	A_j = \int_{(j-1)\xi}^{j\xi} V^\dagger(t_p) \sigma_x V(t_p) dt_p,  B_j = \int_{(j-1)\xi}^{j\xi} V^\dagger(t_p) \sigma_y V(t_p) dt_p, C_j = \int_{(j-1)\xi}^{j\xi} V^\dagger(t_p) \sigma_z V(t_p) dt_p.
\end{equation}
The condition for CUDD sequence in Eq.~(\ref{cu1d}) can then be rewritten as
\begin{align}
I_x &= \sum_k R_z^\dagger(\phi_k) \left[ \cos\phi_k  A_{j(k)} - \sin\phi_k  B_{j(k)} \right] R_z(\phi_k) = 0, \\
I_y &= \sum_k R_z^\dagger(\phi_k) \left[ \sin\phi_k  A_{j(k)} + \cos\phi_k  B_{j(k)} \right] R_z(\phi_k) = 0, \\
I_z &= \sum_k R_z^\dagger(\phi_k)  C_{j(k)}  R_z(\phi_k) = 0.
\end{align}
As $I_\mu$ is traceless Hermitian, the above conditions further reduce to $I_\mu^{1,1}=I_\mu^{1,2}=0$ (with superscripts denoting row and column indices, respectively), and explicitly
\begin{align}
I_x^{1,1} &= \sum_k \left[ \cos\phi_k  a_{j(k)} - \sin\phi_k  b_{j(k)} \right] = 0,  I_y^{1,1} = \sum_k \left[ \sin\phi_k  a_{j(k)} + \cos\phi_k  b_{j(k)} \right] = 0,  \label{eq20}\\
I_z^{1,1} &= \sum_k c_{j(k)} = M \sum_j c_j = 0, \\
I_x^{1,2} &= \sum_k \left[ \cos\phi_k f_{j(k)} - \sin\phi_k g_{j(k)} \right]e^{i\phi_k} = 0,  I_y^{1,2} = \sum_k \left[ \sin\phi_k  f_{j(k)} + \cos\phi_k  g_{j(k)} \right]e^{i\phi_k} = 0, \label{eq22} \\
I_z^{1,2} &= \sum_k h_{j(k)}e^{i\phi_k}  = 0,  \label{eq23} 
\end{align}
where we have defined $a_j=A_j^{1,1},b_j=B_j^{1,1},c_j=C_j^{1,1}, f_j=A_j^{1,2},g_j=B_j^{1,2},h_j=C_j^{1,2}$. We further define  $d_j=a_j + i b_j$, thus Eq. (\ref{eq20}) can be expressed as 
\begin{equation}\label{eq24}
	I_x^{1,1}+ i I_y^{1,1}=\sum_k [ a_{j(k)}+ i b_{j(k)}]e^{i\phi_k}=  \sum_k d_{j(k)} e^{i\phi_k}=0.
\end{equation}
Similarly, we define $q_j=f_j +i g_j$, thus  Eq. (\ref{eq22}) can be expressed as
\begin{equation}\label{eq25}
	I_x^{1,2}+ i I_y^{1,2}=\sum_k [ f_{j(k)}+ i g_{j(k)}]e^{2 i\phi_k}=  \sum_k q_{j(k)} e^{2i\phi_k}=0.
\end{equation}
Overall, the conditions can be summarized as 
\begin{equation}
\begin{cases}\label{eq31}
\displaystyle \sum_k d_{j(k)} e^{i\phi_k}=0, 
& d_j = A_j^{1,1} + i B_j^{1,1}, \\[6pt]

\displaystyle \sum_k q_{j(k)} e^{2i\phi_k}=0, 
& q_j = A_j^{1,2} + i B_j^{1,2}, \\[6pt]

\displaystyle \sum_k h_{j(k)} e^{i\phi_k}=0, 
& h_j = C_j^{1,2}.
\end{cases}
\end{equation}
In addition, we note that the original CUD pulse satisfies  $\sum_{j=1}^4 A_j= \sum_{j=1}^4 B_j= \sum_{j=1}^4 C_j =0$ under the above definitions, which can be equivalently expressed as 
\begin{equation}
	\sum_{j=1}^4 d_j =0,  \sum_{j=1}^4 q_j =0,  \sum_{j=1}^4 c_j =0, \sum_{j=1}^4 h_j =0.
\end{equation}

Next, we further simplify the above conditions by substituting the explicit form of $V(t_p)=e^{-i \Omega t_p \sigma_z/2} e^{-i \Omega t_p \sigma_{\phi_0}}$  into Eq.~(\ref{eqabc}). Note that for angles $\theta_1$ and $\theta_2$ in the $x$-$y$ plane, the Pauli operators satisfy  $[\sigma_{\theta_1},\sigma_{\theta_2}]=2 i \sin(\theta_2-\theta_1)\sigma_z$ and $\sigma_{\phi_0}\sigma_{\theta_1}\sigma_{\phi_0}=\sigma_{2\phi_0-\theta_1}$. In addition, $[\sigma_z,\sigma_{\phi_0}]= 2i \sigma_{\phi_0+\pi/2}$  and  $\sigma_{\phi_0}  \sigma_{z} \sigma_{\phi_0}=-\sigma_z$. Using these relations, we obtain
\begin{align*}
	V^\dag (t_p) \sigma_x V(t_p)
	&=  e^{i \Omega t_p \sigma_{\phi_0}} e^{i \Omega t_p \sigma_z/2} \sigma_x e^{-i \Omega t_p \sigma_z/2} e^{-i \Omega t_p \sigma_{\phi_0}}= e^{i \Omega t_p \sigma_{\phi_0}} \sigma_{-\Omega t_p}  e^{-i \Omega t_p \sigma_{\phi_0}} \\ 
	&=\cos^2(\Omega t_p) \sigma_{-\Omega t_p} +\sin^2(\Omega t_p)  \sigma_{\phi_0}  \sigma_{-\Omega t_p} \sigma_{\phi_0}-i \sin(\Omega t_p) \cos(\Omega t_p)[\sigma_{-\Omega t_p}, \sigma_{\phi_0}] \\
	&= \cos^2(\Omega t_p ) \sigma_{-\Omega t_p } +\sin^2(\Omega t_p)  \sigma_{\Omega t_p +2\phi_0}+\sin(2\Omega t_p )\sin(\Omega t_p +\phi_0) \sigma_z, \\
	V^\dag (t_p) \sigma_y V(t_p)
	&=  e^{i \Omega t_p \sigma_{\phi_0}} e^{i \Omega t_p \sigma_z/2} \sigma_y e^{-i \Omega t_p \sigma_z/2} e^{-i \Omega t_p \sigma_{\phi_0}}= e^{i \Omega t_p \sigma_{\phi_0}} \sigma_{\pi/2-\Omega t_p}  e^{-i \Omega t_p \sigma_{\phi_0}} \\
	&= \cos^2(\Omega t_p) \sigma_{\pi/2-\Omega t_p} +\sin^2(\Omega t_p)  \sigma_{\phi_0}  \sigma_{\pi/2-\Omega t_p} \sigma_{\phi_0}-i \sin(\Omega t_p) \cos(\Omega t_p)[\sigma_{\pi/2-\Omega t_p}, \sigma_{\phi_0}] \\
	&= \cos^2(\Omega t_p) \sigma_{\pi/2-\Omega t_p} +\sin^2(\Omega t_p) \sigma_{\Omega t_p+2\phi_0-\pi/2} -  \sin(2\Omega t_p)\cos(\Omega t_p+\phi_0) \sigma_z, \\
	V^\dag (t_p) \sigma_z V(t_p)
	&=e^{i \Omega t_p \sigma_{\phi_0}} e^{i \Omega t_p \sigma_z/2} \sigma_z e^{-i \Omega t_p \sigma_z/2} e^{-i \Omega t_p \sigma_{\phi_0}}=e^{i \Omega t_p \sigma_{\phi_0}} \sigma_{z}  e^{-i \Omega t_p \sigma_{\phi_0}} \\
	&=\cos^2(\Omega t_p) \sigma_{z} +\sin^2(\Omega t_p)  \sigma_{\phi_0}  \sigma_{z} \sigma_{\phi_0}-i \sin(\Omega t_p) \cos(\Omega t_p)[\sigma_{z}, \sigma_{\phi_0}] \\
	&=\cos(2\Omega t_p) \sigma_{z} +\sin(2\Omega t_p) \sigma_{\phi_0+\pi/2}.
\end{align*} 
This leads to 
\begin{align*}
	[V^\dag (t_p) \sigma_x V(t_p)]^{1,1} &= \sin(2\Omega t_p)\sin(\Omega t_p+\phi_0), [V^\dag (t_p) \sigma_x V(t_p)]^{1,2}=e^{i\Omega t_p}\cos^2(\Omega t_p) + e^{-i(\Omega t_p+2\phi_0)}\sin^2(\Omega t_p), \\ 
	[V^\dag (t_p) \sigma_y V(t_p)]^{1,1} &= -  \sin(2\Omega t_p)\cos(\Omega t_p+\phi_0),  [V^\dag (t_p) \sigma_y V(t_p)]^{1,2}= - i e^{i\Omega t_p}\cos^2(\Omega t_p)+ i e^{-i(\Omega t_p+2\phi_0)}\sin^2(\Omega t_p), \\
	[V^\dag (t_p) \sigma_z V(t_p)]^{1,1} &= \cos(2\Omega t_p), [V^\dag (t_p) \sigma_z V(t_p)]^{1,2} = - i e ^{-i\phi_0 } \sin(2\Omega t_p), 
\end{align*}
and it can be readily verified that $[V^\dag (t_p) \sigma_\mu  V(t_p)]^{2,2}=-[V^\dag (t_p) \sigma_\mu V(t_p)]^{1,1}, [V^\dag (t_p) \sigma_\mu V(t_p)]^{2,1}=[V^\dag (t_p) \sigma_\mu V(t_p)]^{1,2}$.
Therefore, $d_j,q_j$ and $h_j$ in Eq. (\ref{eq31}) can be explicitly calculated by  denoting $\zeta=\Omega t_p $ ($\xi=\pi/(2\Omega)$) 
\begin{align}
	d_j &= \int_{(j-1)\xi}^{j\xi} \left[A_j^{1,1} + i B_j^{1,1}\right] dt_p =  \int_{(j-1)\xi}^{j\xi}  \sin(2\Omega t_p ) e^{i(\Omega t_p +\phi_0-\pi/2)} dt_p  = e^{i(\phi_0-\pi/2)}\int_{(j-1)\pi/2}^{j\pi/2} \frac{e^{2i\zeta}-e^{-2i\zeta}}{2 i \Omega } e^{i\zeta} d\zeta\nonumber\\
	&=\frac{e^{i(\phi_0-\pi/2)}}{2i\Omega} \int_{(j-1)\pi/2}^{j\pi/2} (e^{3i\zeta}-e^{-i\zeta})d\zeta =-\frac{e^{i(\phi_0-\pi/2)}}{2\Omega}\left[\frac{e^{3i\zeta}}{3}+e^{-i\zeta}\right] \biggr|_{(j-1)\pi/2}^{j\pi/2}= \frac{2\sqrt{2}}{3\Omega} e^{i(\phi_0 - \pi/4)} (-i)^{j-1}, \\
	q_j &= \int_{(j-1)\xi}^{j\xi} \left[A_j^{1,2} + i B_j^{1,2}\right] dt_p  =  \int_{(j-1)\xi}^{j\xi}  2\cos^2(\Omega t_p ) e^{i\Omega t_p } dt_p  = \frac{1}{\Omega}  \int_{(j-1)\pi/2}^{j\pi/2} \left[e^{i\zeta}+\frac{e^{3i\zeta}+e^{-i\zeta}}{2} \right] d\zeta
	  \nonumber\\
	  &= \frac{1}{i \Omega}  \left[e^{i\zeta}+\frac{e^{3i\zeta}}{6}-\frac{e^{-i\zeta}}{2}\right] \biggr|_{(j-1)\pi/2}^{j\pi/2}= 
\begin{cases}
 (4+2i)/(3\Omega), & j=1 \\
 (-4+2i)/(3\Omega), & j=2 \\
(-4-2i)/(3\Omega), & j=3 \\
(4-2i)/(3\Omega), & j=4
\end{cases},  \\
	h_j &= \int_{(j-1)\xi}^{j\xi} C_j^{1,2}dt_p =  -i e^{-i\phi_0}\int_{(j-1)\xi}^{j\xi} \sin(2\Omega t_p ) dt_p =  \frac{-i e^{-i\phi_0}}{\Omega} \int_{(j-1)\pi/2}^{j\pi/2} \sin (2\zeta) d\zeta=\frac{-i e^{-i\phi_0}}{\Omega} (-1)^{j-1}.
\end{align}
Using these results and noting that $d_j=d_1 \cdot (-i)^{j-1}, q_3=-q_1,q_4=-q_2,h_1=h_3=-h_2=-h_4$, the conditions in Eq. (\ref{eq31}) can be expressed as 
\begin{equation}
\sum_{m=1}^M A
\begin{pmatrix}
e^{i\phi_{4m-3}} \\
e^{i\phi_{4m-2}}\\
e^{i\phi_{4m-1}} \\
e^{i\phi_{4m}}
\end{pmatrix}
= \mathbf{0}, \quad
A=\begin{pmatrix}
1 & -i & -1 & i \\
2 + i & -2 + i & -(2 + i) & 2 - i \\
1 & -1 & 1 & -1
\end{pmatrix}.
\end{equation}

\subsection{Conditions for maintaining the robustness to Rabi error}
We now derive the fidelity of the CUDD sequence and determine the phase conditions required to preserve robustness against Rabi errors.  For brevity, we denote the evolution operator of the $k$th pulse as
\begin{equation}
W_k^{\phi_k}=
\begin{bmatrix}
\sqrt{1-p} & -i\sqrt{p}\,e^{-i\Phi_k}\\[4pt]
-i\sqrt{p}\,e^{i\Phi_k} & \sqrt{1-p}
\end{bmatrix}
\equiv  c \mathbb{I} -i s \sigma_{\Phi_k},
\label{eq:Wk}
\end{equation}
where 
\begin{equation*}
c\equiv \sqrt{1-p}=\cos \left[\frac{\pi}{2}(1+\varepsilon)\right], 
s \equiv \sqrt{p}=\sin \left[\frac{\pi}{2}(1+\varepsilon)\right],  \sigma_{\Phi_k}=\cos\Phi_k \sigma_x+ \sin\Phi_k \sigma_y
\end{equation*}
with $c^2+s^2=1$ and the single-segment transition probability $p=\sin^2[\tfrac{\pi}{2}(1+\varepsilon)]$. For the ideal case $\varepsilon=0$ one has $c=0$, $s=1$, i.e. each segment is a perfect $\pi$ rotation; we therefore treat $q\equiv 1-p=c^2=\sin^2(\pi\varepsilon/2)\simeq \pi^2\varepsilon^2/4$ as the small expansion parameter.

\subsubsection{Sequence propagator and its trace}
The total sequence evolution operator is  $ U_n =\prod_{k=1}^{n} W_k^{\phi_k}$, where $n=4M$. Its trace can be obtained by  choosing, for each segment, either the term $c I $ or $-i\sqrt{p} \sigma_{\Phi_k}$, which gives
\begin{equation}\label{eq35}
\mathrm{Tr} (U_n)
=\sum_{S\subseteq \{1,\dots,n\}}
c^{|S|}(-i\sqrt{p})^{\,n-|S|}\,
\mathrm{Tr}\left[\widehat{\sigma}(S^{c})\right],
\end{equation}
where $S$ is the set of segments assigned $c \mathbb{I}$, and
$\widehat{\sigma}(S^{c})$ is the ordered product of $\sigma_{\Phi_k}$ over the complementary set, taken in decreasing order of the index.

To proceed, we use the identities $\sigma_\alpha \sigma_\beta= \cos(\alpha-\beta)\mathbb{I} -i \sin(\alpha-\beta)\sigma_z$ and $\sigma_z \sigma_{\alpha}=i \sigma_{\alpha+\pi/2}$, from which we obtain
\begin{equation*}
\mathrm{Tr}\left(\sigma_{\alpha_1}\sigma_{\alpha_2}\cdots \sigma_{\alpha_{2m}}\right)
=2\cos \left(\alpha_1-\alpha_2+\alpha_3-\cdots-\alpha_{2m}\right).
\end{equation*}
For an odd number of factors, the product is purely transverse (a linear combination of $\sigma_x$ and $\sigma_y$), and therefore its trace vanishes.
Thus, Eq. (\ref{eq35}) can be simplified to 
\begin{equation}\label{eq37}
	\frac{\mathrm{Tr} (U_n)}{2}=\sum_{m=0}^{n/2} (-1)^m c^{n-2m} p^m R_{2m} 
\end{equation}
with the geometric coefficients ($\Phi_k=\phi_0+\phi_k$)
\begin{equation*}
R_{2m}\equiv
\sum_{\substack{S^c\subseteq\{1,\ldots,n\}\\|S^c|=2m}}
\cos\left( \Phi_{i_1}-\Phi_{i_2}+\Phi_{i_3}-\cdots-\Phi_{i_{2m}}\right)
=\sum_{\substack{S^c\subseteq\{1,\ldots,n\}\\|S^c|=2m}}
\cos\left( \phi_{i_1}-\phi_{i_2}+\phi_{i_3}-\cdots-\phi_{i_{2m}}\right) ,
\end{equation*}
where $i_1>i_2>\cdots>i_{2m}$ are the ordered elements of the subset $S^c$. In particular $R_0=1$, $R_2=\sum_{i<j}\cos(\phi_j-\phi_i)$. For even $n$, the highest-order coefficient is
$R_n=\cos G$ with
\begin{equation*}
G\equiv\sum_{k=1}^{n/2}\big(\phi_{2k}-\phi_{2k-1}\big).
\end{equation*}

\subsubsection{Perturbative expansion at $p=1$}
For $n=4M$ we have $c^{n-2m}=(c^2)^{2M-m}=(1-p)^{2M-m}$. Reindexing
Eq.~(\ref{eq37}) with $\ell=2M-m$ gives
\begin{equation}
\frac{\mathrm{Tr} (U_{4M})}{2}
=\sum_{\ell=0}^{2M}(-1)^{\ell} (1-p)^{\ell} p^{ 2M-\ell}R_{4M-2\ell}.
\end{equation}
Since $q=1-p$ is of order $\varepsilon^2$, the leading and next-to-leading contributions are
\begin{equation}\label{eq41}
\frac{\mathrm{Tr} (U_{4M})}{2}
=R_{4M}-q\big(2M R_{4M}+R_{4M-2}\big)+ \mathcal{O}(q^{2}).
\end{equation}

\subsubsection{Sequence fidelity and  phase conditions}
Taking the target sequence as the ideal evolution
$U_{\mathrm{target}}=\pm \mathbb{I}$, the sequence fidelity is given by
\begin{equation}
F=\frac{1}{4}\big|\mathrm{Tr}\big(U_{\mathrm{target}}^{\dagger}U_{4M}\big)\big|^{2}= \frac{1}{4}\big|\mathrm{Tr}\big(U_{4M}\big)\big|^{2}.
\end{equation}
At $\varepsilon=0$, only the $\ell=0$ term in Eq.~(\ref{eq41}) contributes, giving $\mathrm{Tr}(U_{4M}/2)\big|_{\varepsilon=0} = R_{4M} = \cos G$, where $G\equiv\sum_{j=1}^{2M}(\phi_{2j}-\phi_{2j-1})$. To ensure $F|_{\varepsilon=0}=1$, we impose
\begin{equation}
\cos G = \pm 1, \quad\text{i.e.}\quad
G=\sum_{j=1}^{2M}(\phi_{2j}-\phi_{2j-1}) = n\pi\ (n\in\mathbb{Z}).
\end{equation}
If we denote $\tilde\phi_k=\sum_{j=1}^k(-1)^{j-1}\phi_j$, $\tilde\phi_0=0$, then the above condition can be expressed as 
\begin{equation}
	e^{i\tilde\phi_{4M}}=\pm 1.
\end{equation}

At $\varepsilon \neq 0$, substituting the above condition $\eta\equiv R_{4M} = \cos G=\pm 1$ into Eq.~(\ref{eq41}), the sequence fidelity becomes 
\begin{equation*}
F = \big[\eta - q (2M\eta + R_{4M-2}) + \mathcal{O}(q^2)\big]^2
  = 1 - 2q\,(2M + \eta\,R_{4M-2}) + \mathcal{O}(q^2),
\end{equation*}
so the leading-order infidelity is
\begin{equation}\label{inf41}
1-F \simeq \frac{\pi^2\varepsilon^2}{2} (2M + \eta R_{4M-2}).
\end{equation}
The fidelity is maximized when
\begin{equation}\label{eq42}
 2M+ \eta R_{4M-2}=0,
\end{equation}
which eliminates the $\mathcal{O}(\varepsilon^2)$ term and pushes the residual infidelity to $\mathcal{O}(\varepsilon^4)$.

\subsubsection{An equivalent and simple geometric constraint to Eq. (\ref{eq42})}
We now introduce a constraint that is both equivalent to Eq. (\ref{eq42}) and geometrically straightforward.
First, we express each segment propagator as
\begin{equation*}
W_k^{\phi_k} = W_k^{\phi_k(0)} D_k,
\end{equation*}
where $W_k^{\phi_k(0)} = -i\sigma_{\Phi_k}$ is the ideal ($\varepsilon=0$)
propagator and
\begin{equation*}
D_k \equiv (W_k^{\phi_k(0)})^\dagger W^{\phi_k}_k
= (i\sigma_{\Phi_k})(c \mathbb{I} - i s \sigma_{\Phi_k})
= s \mathbb{I} + ic\sigma_{\Phi_k}.
\end{equation*}
Since $\sigma_{\Phi_k}^2=\mathbb{I}$, the exponential
$e^{ic \sigma_{\Phi_k}}=\cos c \mathbb{I}+i\sin c \sigma_{\Phi_k}
=(1-\tfrac{c^2}{2})\mathbb{I}+ic\,\sigma_{\Phi_k}+\mathcal{O}(c^3)$.
Using $s=\sqrt{1-c^2}=1-\tfrac{c^2}{2}+\mathcal{O}(c^4)$, this matches $D_k=s \mathbb{I} +ic \sigma_{\Phi_k}$ to second order, so
\begin{equation}
D_k=s \mathbb{I} + ic\sigma_{\Phi_k}= e^{ic \sigma_{\Phi_k}}+\mathcal{O}(c^3),
\end{equation}
a small rotation about $\sigma_{\Phi_k}$ by angle $c=\sqrt{q}\propto\varepsilon$.

Second, we move all error operators $D_k$ to the right of the ideal propagators. Inserting $W_k^{\phi_k}=W_k^{\phi_k(0)}D_k$ into $U_{4M}=W^{\phi_k}_{4M}\cdots W^{\phi_k}_1$ and repeatedly using $D_k W_{k-1}^{\phi_k(0)}\cdots W_1^{\phi_k(0)} =(W_{k-1}^{\phi_k(0)}\cdots W_1^{\phi_k(0)}) \big[Q_{k-1}^\dagger D_k Q_{k-1}\big]$ with $Q_{k-1}=W_{k-1}^{\phi_k(0)}\cdots \\ W_1^{\phi_k(0)}$, we obtain
\begin{equation}
U_{4M}=U_{4M}^{(0)}\,\tilde D_{4M}\cdots\tilde D_1,
\qquad U_{4M}^{(0)}=Q_{4M},
\end{equation}
where each error operator in the toggling frame is
\begin{equation}
\tilde D_k=Q_{k-1}^\dagger D_k Q_{k-1}
=s \mathbb{I} +ic \big(Q_{k-1}^\dagger\sigma_{\Phi_k}Q_{k-1}\big)
\equiv s \mathbb{I}+ic \sigma_{\theta_k}=e^{ic \sigma_{\theta_k}}.
\end{equation}
The conjugated axis $\sigma_{\theta_k}=Q_{k-1}^\dagger\sigma_{\Phi_k}Q_{k-1}$ remains a
transverse Pauli operator. To evaluate $\theta_k$, note that $Q_{k-1}^\dagger(\cdot)Q_{k-1}$ applies the conjugation by $(W_1^{\phi_k(0)})^\dagger\cdots(W_{k-1}^{\phi_k(0)})^\dagger (\cdot)W_{k-1}^{\phi_k(0)} \cdots W_1^{\phi_k(0)}$, i.e.,  successive reflections starting from $\Phi_{k-1}$ (innermost) out
to $\Phi_1$ (outermost). Each reflection maps
$\sigma_\alpha\to\sigma_{2\Phi_j-\alpha}$. Iterating:
\begin{align*}
&\text{after }\Phi_{k-1}:\quad \Phi_k\;\longrightarrow\;2\Phi_{k-1}-\Phi_k,
\nonumber\\
&\text{after }\Phi_{k-2}:\quad\longrightarrow\;
2\Phi_{k-2}-(2\Phi_{k-1}-\Phi_k)=2\Phi_{k-2}-2\Phi_{k-1}+\Phi_k,
\nonumber\\
&\text{after }\Phi_{k-3}:\quad\longrightarrow\;
2\Phi_{k-3}-2\Phi_{k-2}+2\Phi_{k-1}-\Phi_k,
\nonumber\\
&\qquad\vdots\nonumber\\
&\text{after }\Phi_1:\quad\longrightarrow\;
2\Phi_1-2\Phi_2+2\Phi_3-\cdots\pm 2\Phi_{k-1}\mp\Phi_k.
\end{align*}
Defining $T_k=\sum_{j=1}^{k}(-1)^{j-1}\Phi_j$ with $T_0=0$, one can
verify by induction that after all $k{-}1$ reflections the result is
\begin{equation}
\theta_k=T_k+T_{k-1},  \quad T_k=\sum_{j=1}^{k}(-1)^{j-1}\Phi_j.
\end{equation}

Third, we expand each $\tilde D_k=e^{ic\sigma_{\theta_k}}=I+ic\sigma_{\theta_k}+\mathcal{O}(c^2)$ and keep first order in $c$,
\begin{equation}
\prod_{k=1}^{4M}\tilde D_k = \mathbb{I} +ic\sum_{k=1}^{4M}\sigma_{\theta_k}
+\mathcal{O}(c^2).
\end{equation}
The summed generator is a single in-plane Pauli operator,
\begin{equation*}
\sum_{k=1}^{4M}\sigma_{\theta_k}
=\Big(\sum_k\cos\theta_k\Big)\sigma_x+\Big(\sum_k\sin\theta_k\Big)\sigma_y,
\end{equation*}
whose norm is $|\mathcal{S}|$ with $\mathcal{S}\equiv\sum_k e^{i\theta_k}$,
since $(\sum_k\cos\theta_k)^2+(\sum_k\sin\theta_k)^2=|\mathcal{S}|^2$.
Writing $\sum_k\sigma_{\theta_k}=|\mathcal{S}|\,\hat n\cdot\boldsymbol\sigma$
with $\hat n$ a unit in-plane vector, the product re-exponentiates as
\begin{equation*}
\prod_{k=1}^{4M}\tilde D_k
= \mathbb{I}+ic|\mathcal{S}| \hat n\cdot\boldsymbol\sigma+\mathcal{O}(c^2)
= e^{ic|\mathcal{S}| \hat n\cdot\boldsymbol\sigma}+\mathcal{O}(c^2).
\end{equation*}
Thus, we easily get
\begin{equation*}
\mathrm{Tr}\prod_k\tilde D_k = 2\cos(c|\mathcal{S}|).
\end{equation*}

Finally, we note that $U_{\mathrm{target}}=U_{4M}^{(0)}$, so the sequence fidelity can be expressed as 
\begin{equation}
F=\frac{1}{4}\Big|\mathrm{Tr}\big(U_{\mathrm{target}}^\dagger U_{4M}\big)\Big|^2
=\frac14\Big|\mathrm{Tr}\prod_k\tilde D_k\Big|^2
=\cos^2(c|\mathcal{S}|).
\end{equation}
Expanding for small $c^2=q\simeq \pi^2\varepsilon^2/4$, we obtain 
\begin{equation}
1-F\simeq c^2|\mathcal{S}|^2
=q\Big|\sum_{k=1}^{4M}e^{i\theta_k}\Big|^2
=\frac{\pi^2\varepsilon^2}{4}\Big|\sum_{k=1}^{4M}e^{i\theta_k}\Big|^2 .
\end{equation}
The infidelity vanishes to leading order if and only if
\begin{equation}\label{eq50}
\sum_{k=1}^{4M}e^{i\theta_k}=0,
\end{equation}
i.e., the $4M$ toggling-frame axes form a closed polygon on the unit circle. Compared to Eq. (\ref{inf41}), we get the following equation
\begin{equation*}
	\Big|\sum_{k=1}^{4M}e^{i\theta_k}\Big|^2 =2(2M+\eta R_{4M-2}),
\end{equation*}
This reveals that Eq. (\ref{eq50}) is equivalent to Eq. (\ref{eq42}). Furthermore, notice that $\Phi_k=\phi_0+\phi_k$, one has $T_k=\phi_0\cdot\sum_{j=1}^k(-1)^{j-1}+\tilde \phi_k$ with $\tilde \phi \equiv\sum_{j=1}^k(-1)^{j-1}\phi_j$. 
Since $S_k+S_{k-1}=1$ for all $k\ge 1$ (where
$S_k=\sum_{j=1}^k(-1)^{j-1}$), one obtains $\theta_k=\phi_0+\tilde \phi_k+\tilde \phi_{k-1}$
uniformly. The common factor $e^{i\phi_0}$ drops out, leaving
\begin{equation}
\sum_{k=1}^{4M}e^{i(\tilde \phi_k+\tilde \phi_{k-1})}=0,\qquad
\tilde \phi_k=\sum_{j=1}^{k}(-1)^{j-1}\phi_j, \tilde \phi_0=0.
\end{equation}

\subsection{Explicit optimal phase solutions}
The fidelity condition derived above suppresses the Rabi error, while the continuous unitary 1-design condition guarantees robustness against arbitrary static noise. The two requirements are independent and the optimal phases must satisfy both simultaneously. Overall, for a robust CUDD sequence consisting of $M$ CUD blocks, each with four pulses of equal duration and totaling $4M$ pulses, the phases $\{\phi_1,\dots,\phi_k, \dots, \phi_{4M}\}$ are required to satisfy the following conditions:
%
\begin{subequations}\label{eq:all-conditions}
\begin{align}
\label{eq:cud}
\textbf{(CUD)}&\quad
\sum_{m=1}^{M} A\,\boldsymbol{z}_m = \mathbf{0}, \\[8pt]
\label{eq:rabi}
\textbf{(Rabi)}&\quad
\sum_{k=1}^{4M} e^{i(\tilde\phi_k+\tilde\phi_{k-1})} = 0,
\qquad e^{i\tilde\phi_{4M}}=\pm 1,
\end{align}
\end{subequations}
where $\boldsymbol{z}_m=(e^{i\phi_{4m-3}},e^{i\phi_{4m-2}},
e^{i\phi_{4m-1}},e^{i\phi_{4m}})^T$,
$\tilde\phi_k=\sum_{j=1}^k(-1)^{j-1}\phi_j$, $\tilde\phi_0=0$, and
\begin{equation*}
A=\begin{pmatrix}
1 & -i & -1 & i \\
2{+}i & -2{+}i & -(2{+}i) & 2{-}i \\
1 & -1 & 1 & -1
\end{pmatrix}.
\end{equation*}

We now construct explicit phase solutions satisfying all three conditions
in Eqs.~\eqref{eq:cud}--\eqref{eq:rabi}. The system has $4M$ free phase
variables and 9 real constraints (6 from the CUD  and 3 from Rabi robustness), leaving $4M-9$ residual degrees of freedom for $M\ge 3$.

\subsubsection{Symmetric block ansatz}

A powerful simplification arises from the symmetric block ansatz: within each block~$m$, the four phases share a common base $\alpha_m$ and differ only by a fixed offset pattern~$\boldsymbol\delta=(\delta_1,\delta_2,\delta_3,\delta_4)$,
\begin{equation}
\phi_{4(m-1)+j}=\alpha_m+\delta_j,\qquad j=1,2,3,4.
\label{eq:ansatz}
\end{equation}
The block vector then factorizes as
$\boldsymbol{z}_m=e^{i\alpha_m}\boldsymbol{d}$, with
$\boldsymbol{d}=(e^{i\delta_1},e^{i\delta_2},e^{i\delta_3},e^{i\delta_4})^T$,
and the CUD condition becomes
\begin{equation}
\sum_{m=1}^{M}A\boldsymbol{z}_m
=(A\boldsymbol{d})\sum_{m=1}^{M}e^{i\alpha_m}=\mathbf{0}.
\label{eq:cud-ansatz}
\end{equation}
Provided $A\boldsymbol{d}\neq\mathbf{0}$, the CUD condition reduces to
the single complex constraint
\begin{equation}
\sum_{m=1}^{M}e^{i\alpha_m}=0.
\label{eq:polygon}
\end{equation}

For the Rabi robustness condition $e^{i\tilde\phi_{4M}}=\pm 1$, we substitute the ansatz $\phi_k=\phi_{4(m-1)+j}=\alpha_m+\delta_j$ into $\tilde\phi_{4M}=\sum_{k=1}^{4M}(-1)^{k-1}\phi_k$ and separate the $\alpha_m$ and $\delta_j$ contributions:
\begin{equation}
\tilde\phi_{4M}
=\sum_{m=1}^{M}\sum_{j=1}^{4}(-1)^{4(m-1)+j-1}(\alpha_m+\delta_j)
=\sum_{m=1}^{M}\sum_{j=1}^{4}(-1)^{j-1}(\alpha_m+\delta_j),
\end{equation}
where the sign $(-1)^{4(m-1)+j-1}=(-1)^{j-1}$ since $4(m-1)$ is always even. We then expand the above equation, the $\alpha_m$ part gives
$\sum_m\alpha_m\sum_{j=1}^4(-1)^{j-1}=0$ (the alternating sum over
$j=1,...,4$ vanishes), leaving
\begin{equation}
\tilde\phi_{4M}=M\sum_{j=1}^{4}(-1)^{j-1}\delta_j .
\end{equation}
Therefore, if the chosen offsets satisfy $\tilde \delta_4\equiv \sum_{j=1}^4 (-1)^{j-1}\delta_j=j'\pi~ (j'\in\mathbb{Z})$, then the condition $e^{i\tilde\phi_{4M}}=\pm 1$ is automatically satisfied.

For the Rabi robustness condition $\sum_{k=1}^{4M}e^{i(\tilde\phi_k+\tilde\phi_{k-1})}=0$, we evaluate $\tilde\phi_k$ within each block. Splitting the global sum $\tilde{\phi}_k = \sum_{l=1}^k (-1)^{l-1}\phi_l$ into complete blocks and a partial block, and applying the assumed condition $\sum_{j=1}^4 (-1)^{j-1}\delta_j = j'\pi ~(j'\in\mathbb{Z})$, we obtain (for $k=4(m-1)+j$, $j=1,\ldots,4$)
\begin{equation}
\tilde\phi_{4(m-1)+j}=\tilde \phi_{4(m-1)} + \tilde \phi_j
=\sum_{m'=1}^{m-1}\sum_{l=1}^{4}(-1)^{l-1}(\alpha_{m'}+\delta_l)
+\sum_{l=1}^{j}(-1)^{l-1}(\alpha_m+\delta_l)
=j'(m-1)\pi + \sum_{l=1}^{j}(-1)^{l-1}(\alpha_m+\delta_l),
\end{equation}
where the first term $\tilde \phi_{4(m-1)}$ is derived  as established above. Separating the
$\alpha_m$ and $\delta_l$ parts,
\begin{equation}
\tilde\phi_{4(m-1)+j}
=j'(m-1)\pi + \alpha_m\sum_{l=1}^{j}(-1)^{l-1}+\sum_{l=1}^{j}(-1)^{l-1}\delta_l
\equiv  j'(m-1)\pi +  \alpha_m \nu_j+ \tilde \delta _j,
\end{equation}
with $\nu_j=(1,0,1,0)$ and $\tilde \delta _j=\sum_{l=1}^j(-1)^{l-1}\delta_l$ for $j=1,2,3,4$. The  phases $\tilde\phi_k+\tilde\phi_{k-1}$ within the block are then
\begin{equation}
\tilde\phi_k+\tilde\phi_{k-1}=2 j'(m-1)\pi + \alpha_m(\nu_j+\nu_{j-1})+\tilde \delta_j+\tilde \delta_{j-1}
= 2 j'(m-1)\pi + \alpha_m+\tilde \delta _j+\tilde \delta_{j-1},
\end{equation}
since $\nu_j+\nu_{j-1}=1$ for all $j\ge 1$ (with $\nu_0=0$). Thus, we get 
\begin{equation}
\sum_{k=1}^{4M}e^{i(\tilde\phi_k+\tilde\phi_{k-1})}
= \sum_{m=1}^M e^{i\alpha_m}\sum_{j=1}^{4}e^{i(\tilde \delta _j+\tilde \delta _{j-1})}.
\end{equation}
Therefore, if the chosen offsets satisfy $\sum_{j=1}^{4}e^{i(\tilde \delta _j+\tilde \delta _{j-1})}=0$, then the condition   $\sum_{k=1}^{4M}e^{i(\tilde\phi_k+\tilde\phi_{k-1})}=0$ is automatically satisfied.

Overall, under the symmetric block ansatz $\phi_{4(m-1)+j} = \alpha_m + \delta_j, m=1,2,...,M, j=1,2,3,4$, the conditions in Eq.~(\ref{eq:all-conditions}) reduce to
\begin{subequations}\label{condition_sim}
\begin{align}
\label{cud_sim}
\textbf{(CUD)}&\quad
\sum_{m=1}^{M}e^{i\alpha_m}=0, \\[8pt]
\label{rabi_sim}
\textbf{(Rabi)}&\quad
\sum_{j=1}^{4}e^{i(\tilde \delta _j+\tilde \delta _{j-1})}=0,
\qquad \sum_{j=1}^4 (-1)^{j-1}\delta_j = j'\pi,
\end{align}
\end{subequations}
where $j' \in\mathbb{Z}$ and $ \tilde \delta _j=\sum_{l=1}^j(-1)^{l-1}\delta_l, \tilde \delta _0=0 $.

\subsubsection{Valid base and offset patterns}
Equation~(\ref{cud_sim}) imposes the condition that $M$ unit vectors in the complex plane form a closed polygon. Notable families include:
\paragraph{Uniform spacing.} $\alpha_m=(m-1)\cdot 2\pi/M$ (sum of $M$th roots of unity).
\paragraph{Pairwise cancellation ($M$ even).}
$\{\alpha_m\}=\{\beta_1,\beta_1{+}\pi,\ldots,\beta_{M/2},\beta_{M/2}{+}\pi\}$
with $\beta_j ~(j=1,2,...,M/2)$ arbitrary.
\paragraph{General.}
Any set of $M$ angles whose unit-vector sum vanishes.

Equation ~\eqref{rabi_sim} is readily satisfied by any offset pattern $\boldsymbol\delta$ that contains exactly two zeros and two $\pi$'s.  Thus, there are $\binom{4}{2}=6$ such patterns:
\begin{equation}
\begin{aligned}
&(0,\pi,\pi,0),\quad (0,\pi,0,\pi),\quad (0,0,\pi,\pi),\\
&(\pi,0,0,\pi),\quad (\pi,0,\pi,0),\quad (\pi,\pi,0,0).
\end{aligned}
\label{eq:patterns}
\end{equation}

\subsubsection{Explicit sequences}
For the offset pattern $(0,\pi,0, \pi)$, the explicit sequence phases corresponding to uniform spacing and pairwise cancellation of the base pattern are listed in Table~\ref{tab:cudd-phases}.

\begin{table}[h]
\centering
\renewcommand{\arraystretch}{1.3}
\begin{tabular}{lll}
\hline\hline
Sequence & Phases $(\phi_k)$ & $\{\alpha_m\}$ \\
\hline
\multicolumn{3}{l}{\emph{Uniform spacing:} $\alpha_m=(m{-}1)\cdot 2\pi/M$} \\[2pt]
CUDD8  & $(0,1,0,1,\;1,0,1,0)\pi$ & $\{0,\pi\}$ \\
CUDD12 & $(0,1,0,1,\;\frac{2}{3},\frac{5}{3},\frac{2}{3},\frac{5}{3},\;\frac{4}{3},\frac{1}{3},\frac{4}{3},\frac{1}{3})\pi$
       & $\{0,\frac{2\pi}{3},\frac{4\pi}{3}\}$ \\
CUDD16 & $(0,1,0,1,\;\frac{1}{2},\frac{3}{2},\frac{1}{2},\frac{3}{2},\;1,0,1,0,\;\frac{3}{2},\frac{1}{2},\frac{3}{2},\frac{1}{2})\pi$
       & $\{0,\frac{\pi}{2},\pi,\frac{3\pi}{2}\}$ \\
CUDD20 & $(0,1,0,1,\;\frac{2}{5},\frac{7}{5},\frac{2}{5},\frac{7}{5},\;\frac{4}{5},\frac{9}{5},\frac{4}{5},\frac{9}{5},\;\frac{6}{5},\frac{1}{5},\frac{6}{5},\frac{1}{5},\;\frac{8}{5},\frac{3}{5},\frac{8}{5},\frac{3}{5})\pi$
       & $\{0,\frac{2\pi}{5},\frac{4\pi}{5},\frac{6\pi}{5},\frac{8\pi}{5}\}$ \\[6pt]
\hline
\multicolumn{3}{l}{\emph{Pairwise cancellation} ($M$ even): $\{\alpha_m\}=\{\beta_1,\beta_1{+}\pi,\ldots,\beta_{M/2},\beta_{M/2}{+}\pi\}$} \\[2pt]
CUDD8  & $(0,1,0,1,\;1,0,1,0)\pi$ & $\{0,\pi\}$ \\
CUDD16 & $(0,1,0,1,\;1,0,1,0,\;\frac{1}{4},\frac{5}{4},\frac{1}{4},\frac{5}{4},\;\frac{5}{4},\frac{1}{4},\frac{5}{4},\frac{1}{4})\pi$
       & $\{0,\pi,\frac{\pi}{4},\frac{5\pi}{4}\}$ \\
CUDD16 & $(0,1,0,1,\;1,0,1,0,\;\frac{1}{2},\frac{3}{2},\frac{1}{2},\frac{3}{2},\;\frac{3}{2},\frac{1}{2},\frac{3}{2},\frac{1}{2})\pi$
       & $\{0,\pi,\frac{\pi}{2},\frac{3\pi}{2}\}$ \\
\hline\hline
\end{tabular}
\caption{Phases of CUDD sequences with $M$ blocks (totaling $4M$ segments), using offset $\boldsymbol\delta=(0,\pi,0,\pi)$. Each phase is defined modulo $2\pi$.}
\label{tab:cudd-phases}
\end{table}

\section{Experimental details}

\subsection{Experimental platform and device details}
Our experiments were performed on the Logical Qubit superconducting quantum cloud platform (https://cloud.logicalqubit.com). The backend used in this work was \texttt{QZ01-surface\_code}. As reported by the platform's device overview, this backend comprises 17 physical qubits and 24 couplers, and supports both circuit-level and pulse-level (\texttt{wave}) task submissions. The cloud platform's pulse-level control enables direct benchmarking of CUDD against conventional DD sequences. All pulse sequences were constructed and submitted through the backend's \texttt{wave} task interface, which allows user-defined waveforms with 1~ns timing resolution. Platform-calibrated parameters (e.g., drive amplitude, frequency, and readout settings) were retrieved from the backend configuration at the start of each session.

The device parameters reported by the platform at the time of data acquisition are summarized as follows. Key median metrics include a single-qubit gate fidelity \(F_{\mathrm{SQ}}\) of \(99.926\%\), a two-qubit gate fidelity \(F_{\mathrm{CZ}}\) of \(99.460\%\), and readout fidelities \(F_{00}\) and \(F_{11}\) of \(0.996\) and \(0.980\), respectively. The median relaxation time is \(T_1 = 36.52\ \mu\mathrm{s}\), the median echo coherence time is \(T_{2\mathrm{E}} = 9.03\ \mu\mathrm{s}\), and the median Ramsey coherence time is \(T_2^* = 2.66\ \mu\mathrm{s}\). The platform also reports a logical error rate of \(5.26 \times 10^{-2}\) for this backend.

\subsection{Calibration procedures and calibrated parameters}

All experiments were executed via the LQCloud \texttt{run\_wave} interface, which enables direct pulse-level control by accepting arbitrary waveforms on both the  $XY$ (microwave) and $Z$ (flux) channels of each qubit, with a sampling resolution of 1~ns. Each data point was averaged over 2000 single-shot measurements to ensure statistical reliability. Prior to every experimental session, the qubit parameters were recalibrated starting from the provider's published baseline values, following the procedures outlined below:

\paragraph*{Step 1: Readout discrimination.}
The qubit was prepared in $|0\rangle$ (no drive) and in $|1\rangle$ (the provider's Gaussian DRAG $\pi$ pulse), and the demodulated single-shot IQ points of both preparations were clustered with a two-component $k$-means classifier. The cluster containing most $|0\rangle$ shots was assigned to $|0\rangle$ and the other to $|1\rangle$. This classifier was used to convert IQ data into state populations in all subsequent measurements.

\paragraph*{Step 2: Qubit frequency calibration.}
The drive frequency was refined with a Ramsey sequence
$\pi/2 - \tau - \pi/2_{(\phi)}$, with $\tau$ swept from 0 to 500~ns in 20~ns steps. The phase of the second $\pi/2$ pulse was advanced as $\phi = 2\pi f_{\mathrm{ramp}}\tau$ with an artificial detuning $f_{\mathrm{ramp}} = 10$~MHz. The measured population was fitted to
\begin{equation}
  P_{1}(\tau) = A\, e^{-\tau/T_2^{*}} \cos\!\left(2\pi f \tau + \phi_0\right) + B ,
\end{equation}
with the initial guess for $f$ taken from a least-squares periodogram. The calibrated frequency of the $|0\rangle \leftrightarrow |1\rangle$ transition was obtained as $f_{10} = f_{\mathrm{drive}} + f_{\mathrm{ramp}} - f$.

\paragraph*{Step 3: Calibration of Gaussian $\pi$ pulses for reference.}
The provider's Gaussian $\pi$ pulse (FWHM 25~ns, total duration 75~ns, with fixed DRAG correction) served as the amplitude reference for the square pulses used in the experiments. Its amplitude was further refined by error amplification at the updated frequency. Ideally, a train of $2N$ $\pi$ pulses returns the qubit to $|0\rangle$; thus, the amplitude maximizing $P(|0\rangle)$ was taken as optimal. We performed a coarse scan ($N=1$, $\pm 0.2$ around the provider's value), followed by a fine scan ($N=4$, $\pm 0.05$ around the coarse optimum). Each scan comprised 10 points, with the peak determined by a Gaussian fit. The optimal amplitude from the $N=4$ scan was adopted. All carrier phases were evaluated on the global time axis to ensure phase continuity across concatenated pulses.

\paragraph*{Step 4: Calibration of  square $\pi$ pulses for the conventional DD.}
The conventional DD sequences (CPMG, XY4, and UR4) were implemented using square $\pi$ and $\pi/2$ pulses of durations 30~ns and 15~ns, respectively, applied along the $X$ or $Y$ axis at a Rabi frequency $\Omega$. The pulse amplitude was calibrated by applying $N=9$ consecutive square $\pi$ pulses, which ideally return the qubit to $|1\rangle$, and scanning 41 values between 0.80 and 1.00 times the calibrated Gaussian $\pi$-pulse amplitude. The amplitude maximizing $P(|1\rangle)$ was selected. We then verified the calibration using single $\pi$ and $\pi/2$ pulses, confirming the expected populations $P(|1\rangle)\approx 1$ and $P(|1\rangle)\approx 0.5$, respectively. 

\paragraph*{Step 5: Calibration of  shaped pulses for CUDD.}
As described in the main text, CUDD is rescaled such that its maximum amplitude does not exceed the Rabi frequency. Accordingly, the transverse components along the $X$ and $Y$ axes have rescaled Rabi frequency $(2/\sqrt{5})\Omega$. CUDD also requires $Z$-axis control, whose amplitude was calibrated using the sequence $(\pi/2)_x-Z(30~\mathrm{ns})-(\pi/2)_x$, with a 30~ns flux pulse inserted between two square $\pi/2$ pulses. The $Z$-channel amplitude was swept from 0 to 0.042 in 51 steps. The first minimum of $P(|1\rangle)$ identifies a $\pi$ rotation about the $Z$ axis, defining the calibrated amplitude $z_{\pi}$. The $Z$ component of the CUDD drive was then set to $u_z=z_{\pi}/(2\pi\sqrt{5})$.

The calibrated parameters of each qubit are listed in Table~\ref{tab:calib}. 

\begin{table}[htbp!]
  \centering
  \small
  \begin{tabular}{@{}p{0.4\linewidth}ccc@{}}
    \toprule
    Parameter & Q3 & Q4 & Q13 \\
    \midrule
    Readout assignment ($k$-means), $P(0|0)$ / $P(1|1)$
      & 0.988 / 0.972 & 0.994 / 0.950 & 0.995 / 0.966 \\
    Readout fidelity, provider, $F_0$ / $F_1$
      & 0.998 / 0.987 & 0.997 / 0.982 & 0.997 / 0.982 \\
    $f_{10}$, provider (MHz)
      & $-40.000$ & $250.000$ & $100.000$ \\
    $f_{10}$, Ramsey-calibrated (MHz)
      & $-40.159(8)$ & $249.424(16)$ & $99.862(19)$ \\
    $T_2^{*}$ from Ramsey fit ($\mu$s)
      & $\approx 9.9$ & $\approx 3.4$ & $\approx 7.1$ \\
    Rabi frequency $\Omega$   & 0.3168 & 0.8711 & 0.6362 \\
    Check: single square $\pi$ / $\pi/2$, $P(|1\rangle)$
      & 0.962 / 0.496 & 0.962 / 0.492 & 0.968 / 0.482 \\
    $Z$ $\pi$ amplitude $z_{\pi}$ (30~ns)
      & 0.02184 & 0.02184 & 0.02100 \\
    CUDD $z$ amplitude $u_z = z_{\pi}/(2\pi\sqrt{5})$
      & $1.554\times10^{-3}$ & $1.554\times10^{-3}$ & $1.495\times10^{-3}$ \\
    \bottomrule
  \end{tabular}
  \caption{Calibrated single-qubit parameters. Numbers in parentheses are fit
  uncertainties ($1\sigma$); ``provider'' denotes the value published by the provider. Amplitudes are given in the dimensionless DAC units of the control electronics.}
  \label{tab:calib}
\end{table}

\begin{figure}
\begin{center}
\includegraphics[width=0.7\linewidth]{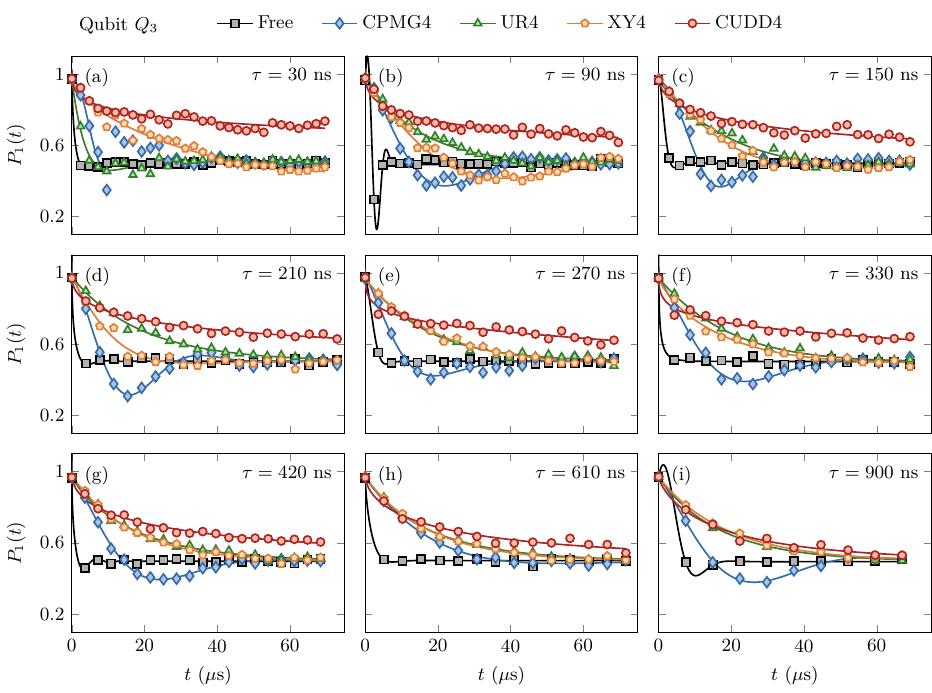}
\end{center}	
\caption{Coherence decay of qubit $Q_3$ under different dynamical decoupling sequences. Each panel shows the probability $P_1(t)$ of measuring the state $|1\rangle$ as a function of the total evolution time $t$, for free evolution (Free) and for the CPMG4, UR4, XY4 and CUDD4 sequences. Panels (a)--(i) correspond to pulse intervals $\tau = 30, 90, 150, 210, 270, 330, 420, 610$ and $900$~ns. Markers are experimental data; solid lines are fits to $P_1(t) = C_\text{inf} + A\cos(\omega t + \Delta)\exp[-(t/T_2)^p]$, where $C_\text{inf}$ is fixed to the long-time plateau of the Free curve at each $\tau$. The fitted parameters are listed in Table~\ref{tab:fit_Q3}.}\label{fig:decay_Q3}
\end{figure}

\begin{figure}
\begin{center}
\includegraphics[width=0.7\linewidth]{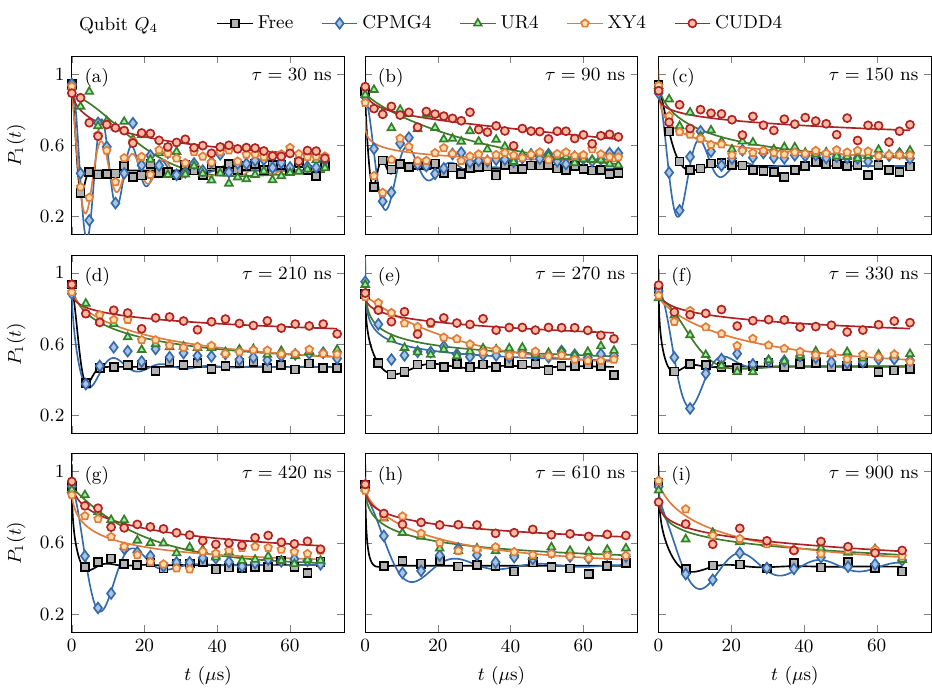}
\end{center}	
\caption{Same as Fig.~\ref{fig:decay_Q3}, for qubit $Q_4$.
The fitted parameters are listed in Table~\ref{tab:fit_Q4}.}\label{fig:decay_Q4}
\end{figure}

\begin{figure}[htbp!]
\begin{center}
\includegraphics[width=0.7\linewidth]{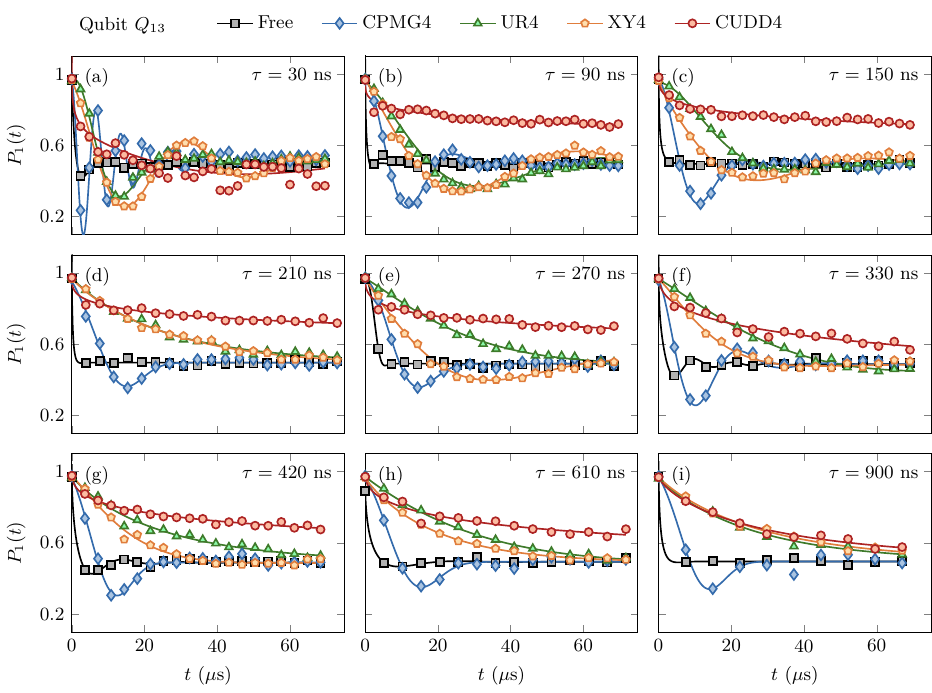}
\end{center}	
\caption{Same as Fig.~\ref{fig:decay_Q3}, for qubit $Q_{13}$.
The fitted parameters are listed in Table~\ref{tab:fit_Q13}.}\label{fig:decay_Q13}
\end{figure}

\begin{table}
  \centering
  \includegraphics[width=\linewidth]{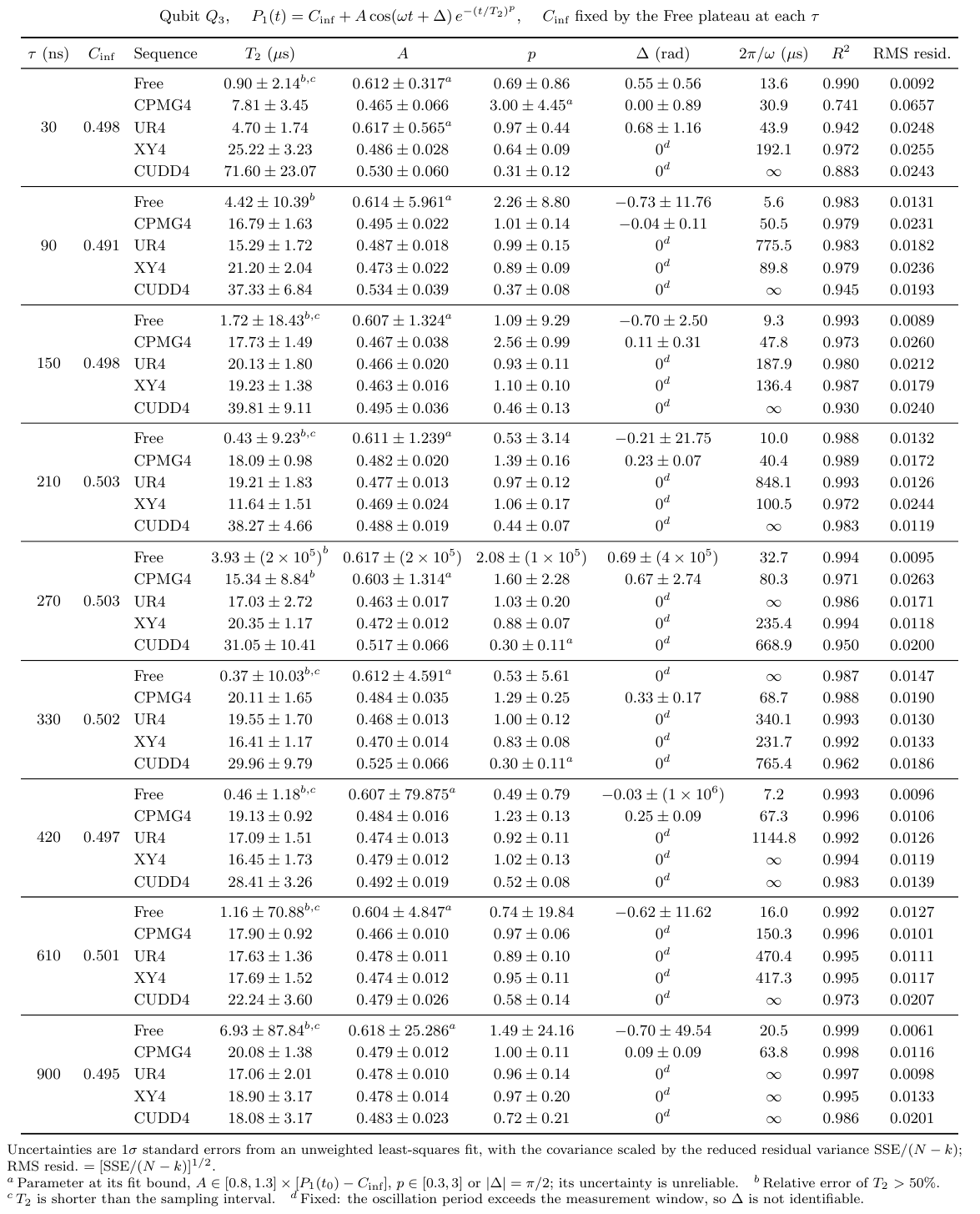}
  \caption{Fitted decay parameters of qubit $Q_3$ for each pulse interval $\tau$ and sequence, obtained from the fits shown in Fig.~\ref{fig:decay_Q3}. The data are fitted to $P_1(t) = C_\text{inf} + A\cos(\omega t + \Delta)\exp[-(t/T_2)^p]$, where $T_2$ is the time for the envelope to decay to $1/e$, $p$ is the stretching exponent, and $2\pi/\omega$ is the oscillation period. At each $\tau$, $C_\text{inf}$ is fixed to the mean of the Free curve at long times and is not a fit parameter. Uncertainties are $1\sigma$ standard errors from an unweighted least-squares fit. $R^2$ and the RMS residual indicate the fit quality. Superscripts mark fits whose parameters should be interpreted with caution (see notes below the table).}
  \label{tab:fit_Q3}
\end{table}

\begin{table}
  \centering
  \includegraphics[width=\linewidth]{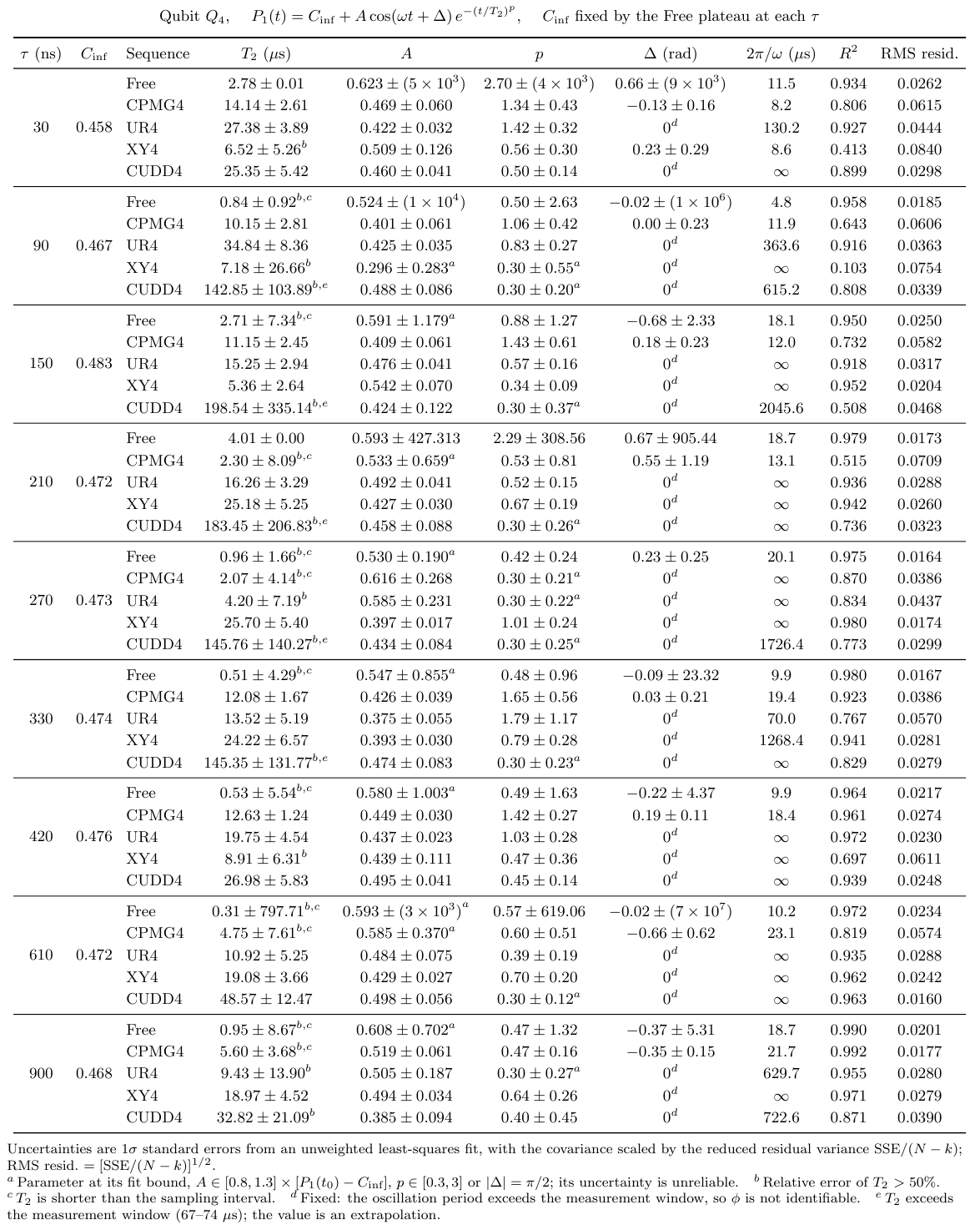}
  \caption{Same as Table~\ref{tab:fit_Q3}, for qubit $Q_4$ (Fig.~\ref{fig:decay_Q4}).}
  \label{tab:fit_Q4}
\end{table}

\begin{table}
  \centering
  \includegraphics[width=\linewidth]{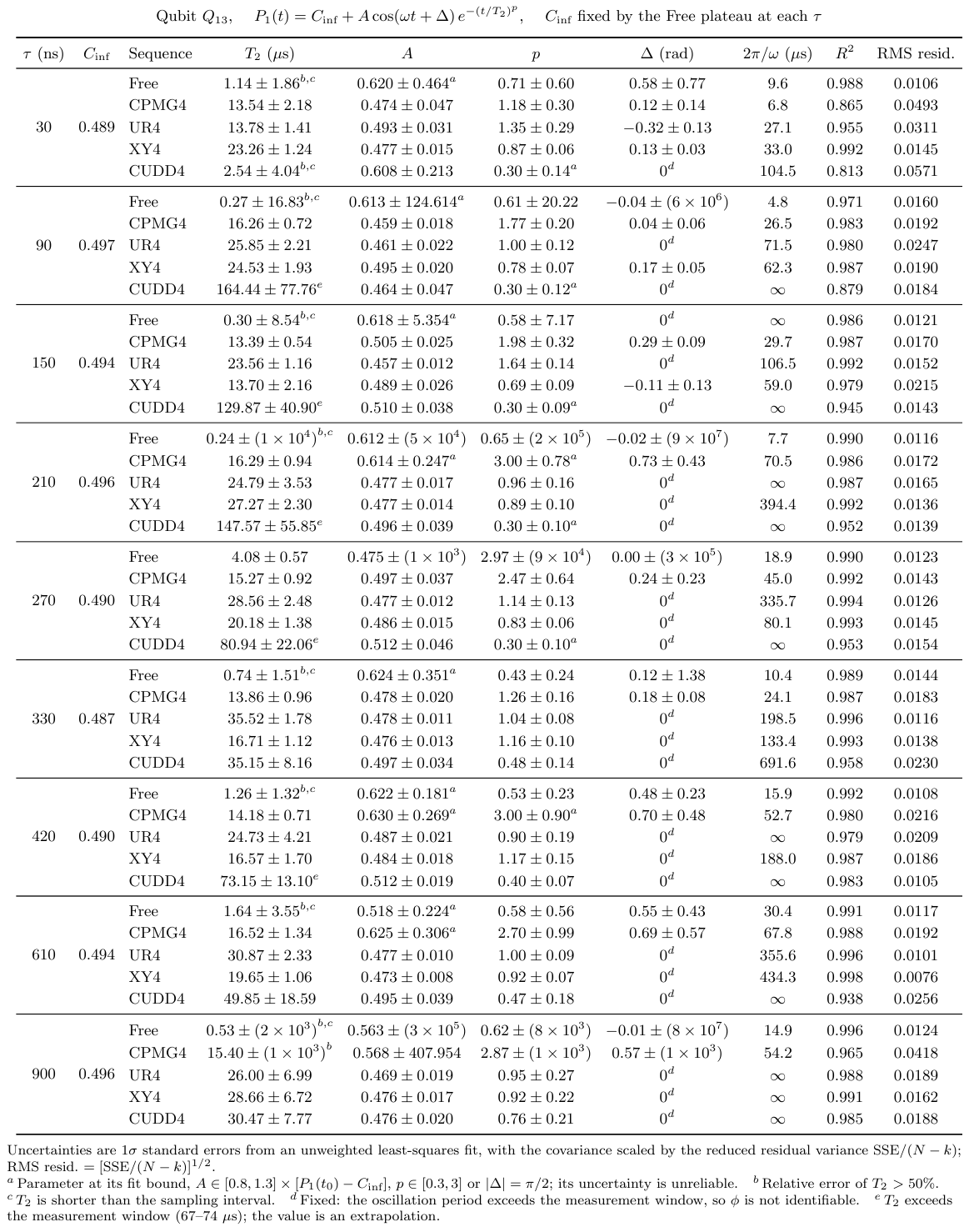}
  \caption{Same as Table~\ref{tab:fit_Q3}, for qubit $Q_{13}$ (Fig.~\ref{fig:decay_Q13}).}
  \label{tab:fit_Q13}
\end{table}

\end{document}